\documentclass{aastex63}
\usepackage{amsmath}
\usepackage{ae,aecompl}
\usepackage{lineno}

\shorttitle{Stabilizing Thin Disks}
\shortauthors{Mishra et al.}

\begin{document}

\title{The Role of Strong Magnetic Fields in Stabilizing Highly Luminous, Thin Disks}

\correspondingauthor{Bhupendra Mishra}
\email{mishra\_b@lanl.gov}

\author{Bhupendra Mishra}
\affiliation{Theoretical Division, Los Alamos National Laboratory, Los Alamos, NM 87545, USA}

\author[0000-0002-5786-186X]{P. Chris Fragile}
\affiliation{Department of Physics and Astronomy, College of Charleston, Charleston, SC 29424, USA}
\affil{Kavli Institute for Theoretical Physics, Kohn Hall, University of California, Santa Barbara, CA 93107, USA}

\author[0000-0002-0028-8054]{Jessica Anderson}
\affiliation{Department of Physics and Astronomy, College of Charleston, Charleston, SC 29424, USA}

\author{Aidan Blankenship}
\affiliation{Department of Physics and Astronomy, College of Charleston, Charleston, SC 29424, USA}

\author{Hui Li}
\affiliation{Theoretical Division, Los Alamos National Laboratory, Los Alamos, NM 87545, USA}

\author{Krzysztof Nalewajko}
\affiliation{Nicolaus Copernicus Astronomical Center, Polish Academy of Sciences, Bartycka 18, 00-716 Warsaw, Poland}

\date{Accepted XXX. Received YYY; in original form ZZZ}

\begin{abstract}
We present and analyze a set of three-dimensional, global, general relativistic radiation magnetohydrodynamic simulations of thin, radiation-pressure-dominated accretion disks surrounding a non-rotating, stellar-mass black hole. The simulations are initialized using the Shakura-Sunyaev model with a mass accretion rate of $\dot{M} = 3 L_\mathrm{Edd}/c^2$ (corresponding to $L=0.17 L_\mathrm{Edd}$). Our previous work demonstrated that such disks are thermally unstable when accretion is driven by an $\alpha$-viscosity. In the present work, we test the hypothesis that strong magnetic fields can both drive accretion through the magneto-rotational instability and restore stability to such disks. We test four initial magnetic field configurations: 1) a zero-net-flux case with a single, radially extended set of magnetic field loops (dipole); 2) a zero-net-flux case with two radially extended sets of magnetic field loops of opposite polarity stacked vertically (quadrupole); 3) a zero-net-flux case with multiple radially concentric rings of alternating polarity (multi-loop); and 4) a net-flux, vertical magnetic field configuration (vertical). In all cases, the fields are initially weak, with the gas-to-magnetic pressure ratio $\gtrsim 100$. Based on the results of these simulations, we find that the dipole and multi-loop configurations remain thermally unstable like their $\alpha$-viscosity counterpart, in our case collapsing vertically on the local thermal timescale and never fully recovering. The vertical case, on the other hand, stabilizes and remains so for the duration of our tests (many thermal timescales). The quadrupole case is intermediate, showing signs of both stability and instability. The key stabilizing factor is the ability of specific field configurations to build up and sustain strong, $P_\mathrm{mag} \gtrsim 0.5P_\mathrm{tot}$, toroidal fields near the midplane of the disk. We discuss the reasons why certain configurations are able to do this effectively and others are not. We then compare our stable simulations to the standard Shakura-Sunyaev disk.
\end{abstract}


\section{Introduction}

From the earliest work on thin $(H/R \ll 1)$ accretion disks based on the $\alpha$-viscosity prescription \citep{Shakura73}, there have been notable problems in region ``A,'' where the vertical pressure support comes from radiation and the opacity is dominated by electron scattering. Region A covers radii $R/r_g \lesssim 600(L/L_\mathrm{Edd})^{16/21}$, where $r_g = GM/c^2$ is the gravitational radius and $L_\mathrm{Edd} = 1.2 \times 10^{38} (M/M_\odot)$ erg s$^{-1}$ is the Eddington luminosity of a black hole of mass $M$. In this region, the disk is predicted to be both thermally \citep{Shakura76} and viscously \citep{Lightman74} unstable. The thermal instability arises because the disk heating rate per unit area, $Q^+$, and cooling rate per unit area, $Q^-$, depend on different powers of the mid-plane pressure for a fixed surface-density, $\Sigma$, as 
\begin{equation}
\left. \frac{\mathrm{d} \ln Q^+}{\mathrm{d} \ln P_{\mathrm{rad},0}} \right\vert_\Sigma = 2
\label{eqn:heating1}
\end{equation}
and 
\begin{equation}
\left. \frac{\mathrm{d} \ln Q^-}{\mathrm{d} \ln P_{\mathrm{rad},0}} \right\vert_\Sigma = 1 ~,
\label{eqn:cooling}
\end{equation}
such that small deviations in $P_{\mathrm{rad},0}$ may lead to runaway heating or cooling. The viscous instability, which typically acts on a longer timescale than the thermal instability, arises due to an inverse correlation between the vertically integrated stress, $W_{R\phi}$, and the surface density, $\Sigma$, which can cause the disk to break up into rings of high and low surface density \citep{Lightman74,Mishra16}. In region ``B'' (gas-pressure-supported, but still scattering dominated), by contrast, which occurs at larger radii or sufficiently low luminosity, the Shakura-Sunyaev solution is predicted to be stable.

Previous shearing box \citep{Jiang13,Ross17} and global simulations \citep{Teresi04,Mishra16,Fragile18b} have largely confirmed the thermal instability of these disks. In \citet{Fragile18b}, multiple $\alpha$-viscosity simulations that started on the radiation-pressure-dominated (region A) branch underwent runaway cooling until they collapsed down to the gas-pressure-dominated (region B) branch of the thermal equilibrium (S-) curve. Only simulations starting on that lower branch remained stable for more than a thermal timescale. Curiously, we did not see any examples of runaway heating, even in one case where the initial gas temperature was perturbed upward by 50\%. A similar preference toward cooling and collapse was noted in \citet{Teresi04}.

All of this is particularly puzzling in light of observations of black hole X-ray binaries (BHXRBs), for which the spectra look most disk-like (soft, with a prominent thermal bump around 1 keV) and stable (rms variability $\lesssim 3$\%) whenever $L/L_\mathrm{Edd} =$ 0.1-0.2 \citep[e.g.][]{vanderKlis04,Done07}. In other words, there is no sign of the thermal or viscous instabilities previously mentioned, precisely in the luminosity range when such disks are predicted to be unstable. Two possible exceptions are GRS 1915+105 \citep{Belloni97,Neilsen11} and IGR J17091-3624 \citep{Altamirano11,Zhang14}, which show evidence for limit cycle behavior that may be consistent with a thermal instability \citep{Honma91,Szuszkiewicz98}, although this seems to be limited to when those sources are at their highest (possibly super-Eddington) luminosity \citep{Done04}.

One proposed solution to the dilemma of thermal instability has been to invoke strong magnetic fields to provide the additional support needed to stabilize the disk \citep{Begelman07,Oda09,Sadowski16}. This is because, while the cooling rate is insensitive to the magnetic pressure and still follows eq. (\ref{eqn:cooling}), the heating rate is not. Instead, the heating rate in a strongly magnetized disk scales as \citep{Sadowski16}
\begin{equation}
\left. \frac{\mathrm{d} \ln Q^+}{\mathrm{d} \ln P_{\mathrm{rad},0}} \right\vert_{\Sigma,P_{\mathrm{mag},0}} = 2(1-\beta^{-1}_r) ~,
\label{eqn:heating2}
\end{equation}
where $\beta_r = P_{\mathrm{tot},0}/P_{\mathrm{mag},0}$ and we have ignored gas pressure. This allows us to quantify how strong the magnetic field must be, as $\beta_r^{-1} > 0.5$ leads to 
\begin{equation}
\left. \frac{\mathrm{d} \ln Q^-}{\mathrm{d} \ln P_{\mathrm{rad},0}} \right\vert_\Sigma > \left.\frac{\mathrm{d} \ln Q^+}{\mathrm{d} \ln P_{\mathrm{rad},0}} \right\vert_{\Sigma,P_{\mathrm{mag},0}} ~,
\end{equation}
which restores stability. In this work, we set out to test the idea of magnetic stabilization through a set of numerical experiments. 

Each of our simulations starts from the \citet{Shakura73} disk solution with $\dot{M}=3L_\mathrm{Edd}/c^2$, i.e. one on the unstable branch and in the prescribed luminosity range ($L=\eta\dot{M}c^2\approx0.057\dot{M}c^2=0.17L_\mathrm{Edd}$). The question we want to ask is: Can an initially weak magnetic field be amplified self-consistently to the required strength to stabilize these disks in this luminosity range? 

We consider four different seed magnetic field configurations, all with $\beta_{\mathrm{mid,}0}=P_{\mathrm{gas,}0}/P_{\mathrm{mag,}0} \gtrsim 100$ and, hence, $\beta_r^{-1} \ll 0.5$. In one, we consider a single poloidal field anti-node centered far from the black hole, giving a very radially extended dipole field configuration. In a second case, we consider two radially extended poloidal loops of opposite polarity stacked vertically, one above the midplane, one below. Both of these field configurations start out weak, though they will be subject to strong shear amplification from the orbital motion of the disk. Similar radially extended fields were reported in \citet{Sadowski16}, with the dipole field deemed unstable and the quadrupole deemed stable. However, it was not entirely clear from that study why the one configuration was stable while the other was not. Hence, our decision to revisit both. In a third case, the field consists of numerous small poloidal loops of alternating polarity, with length scales comparable to the local disk height, arranged in concentric radial rings. For such a configuration, the field is unlikely to amplify sufficiently to offer significant pressure support, and therefore we expect to see a thermal runaway analogous to our previous simulations \citep{Mishra16}. In the final case, we consider a vertical magnetic field threading through the disk. This field configuration is also subject to shear amplification and can reach strengths sufficient to provide magnetic pressure support, as shown in shearing box \citep{Salvesen16} and global \citet{Mishra20} studies. However, ours is the first three dimensional, global, general relativistic, radiation MHD simulations to consider this configuration. 

We elected not to test a purely toroidal field configuration, as we have already shown in \citet{Fragile17} that even a strong toroidal field, {\em on its own}, is not enough to stabilize an initially radiation-pressure-dominated disk. This is because strong toroidal fields decay on roughly the local orbital timescale due to magnetic buoyancy. However, in cases where a radial or vertical seed field is present, such as in our dipole, quadrupole, and vertical field cases, the toroidal magnetic field can be continually replenished by the $\Omega$-dynamo. As long as this replenishment happens fast enough to keep the field strong, this may be enough to stabilize the disk. Alternatively, if a purely toroidal field is able to generate local net poloidal flux through an $\alpha$-$\Omega$ dynamo, as in the thick disk simulations of \citet{Liska20}, then perhaps such a field configuration could yield a stable disk. We do not explore this possibility.

The remainder of our paper is organized as follows: In Section \ref{sec:setup}, we describe the numerical procedures used in our simulations; in Section \ref{sec:results}, we present evidence regarding the stability of each simulation; in Section \ref{sec:diskprof} we describe the vertical profiles of the disk; in Section \ref{sec:Shakura}, we compare the properties of our stable simulations to the predictions of the Shakura-Sunyaev model; in Section \ref{sec:Sadowski}, we discuss how our results fit in with previous, comparable simulations; and finally, in Section \ref{sec:conclusion}, we present our conclusions. All of our simulations assume a non-spinning black hole of mass $M = 6.62 M_\odot$; therefore, our distance unit, $GM/c^2$, is equal to 9.8 km, and our time unit, $GM/c^3$ is equal to $3.3\times 10^{-5}$ s.

\section{Numerical Setup}
\label{sec:setup}

As stated previously, the simulations presented here start from a Shakura-Sunyaev disk with $\dot{M}=3L_\mathrm{Edd}/c^2$. For the hydrodynamic and radiation variables, we follow the initialization steps described in \citet{Fragile18b}. In order to initialize the Shakura-Sunyaev solution, we assume the viscosity parameter, $\alpha$, to be 0.02, based on previous similar simulations \citep[e.g.,][]{Mishra16,Sadowski16} and an adiabatic equation of state with $\gamma=5/3$. However, we emphasize that the current simulations do not employ any form of explicit viscosity. On top of this disk we impose various initially weak ($\beta_{\mathrm{mid},0} \gtrsim 100$) seed magnetic field configurations. We choose to start with weak magnetic fields to test whether the accretion process itself can amplify them to the required strengths.

All simulations are carried out using the general relativistic, radiation, magnetohydrodynamics (GRRMHD) code, {\it Cosmos++} \citep{Anninos05}. We use the high resolution shock capturing (HRSC) scheme described in \citet{Fragile12} to solve for the flux and gravitational source term of the gas and radiation. Rather than evolving the magnetic fields directly, we instead evolve the vector potential and recover the fields from it as needed, as described in \citet{Fragile19}.

For the radiation, we use the $\textbf{M}_{\textbf{1}}$ closure scheme described in \citet{Fragile14}, which retains the first two moments of the radiation intensity and (average) radiative flux. We use grey (frequency-independent) opacities, which are captured in the radiation four-force density (coupling) term:
\begin{eqnarray}
    G^{\mu} & = &-\rho (\kappa^\mathrm{a}_F+\kappa^s)R^{\mu \nu}u_{\nu}- \rho \left\{ \left[\kappa^s+4\kappa^s\left(\frac{T_{\mathrm{gas}}-T_{\mathrm{rad}}}{m_e}\right) +\kappa^\mathrm{a}_F - \kappa^\mathrm{a}_J \right]   R^{\alpha \beta}u_\alpha u_\beta +\kappa^\mathrm{a}_\mathrm{P}a_RT^4_\mathrm{gas}\right\} u^\mu ~,
\label{eqn:Gmu}
\end{eqnarray}
where $\kappa^\mathrm{a}_\mathrm{P}=2.8\times10^{23}\,T^{-7/2}_\mathrm{K} \rho_{\mathrm{cgs}}$ cm$^2$ g$^{-1}$ and $\kappa^\mathrm{a}_\mathrm{R}=7.6\times10^{21}\,T^{-7/2}_\mathrm{K}\,\rho_{\mathrm{cgs}}$ cm$^2$ g$^{-1}$ are the Planck and Rosseland mean opacities for free-free absorption, respectively, $\kappa^s = 0.34$ cm$^2$ g$^{-1}$ is the opacity due to electron scattering, $R^{\mu\nu}$ is the radiation stress tensor, $u^\mu$ is the fluid four-velocity, $T_{\mathrm{K}}$ is the ideal gas temperature of the fluid in Kelvin, and $\rho_{\mathrm{cgs}}$ is density in g cm$^{-3}$. Thus, we are assuming Kramers-type opacity laws, with the Rosseland mean also used for the flux mean, $\kappa^\mathrm{a}_\mathrm{F} = \kappa^\mathrm{a}_\mathrm{R}$, and the Planck mean used for the J-mean, $\kappa^\mathrm{a}_\mathrm{J} = \kappa^\mathrm{a}_\mathrm{P}$. Since the simulations capture turbulence, reconnection, and shock heating as well as radiative cooling directly, we do not need to include any artificially imposed heating or cooling terms.

To invert the conserved fields to the corresponding primitives, we use the {\it Cosmos++} 9D primitive solver, named for the 9 variables that make up the solve. This step uses a Newton-Raphson iterative technique and linear matrix solver to numerically invert the Jacobian matrix composed of derivatives of the conserved fields with respect to the primitive ones. It also simultaneously accomplishes the implicit forward integration of the radiation source term (eq. \ref{eqn:Gmu}) and produces a fully updated set of primitive fields. Details of this procedure are provided in \citet{Fragile14}. In cases where the primitive solver fails to converge or settles on an nonphysical solution, the primitive values of the surrounding zones from the previous timestep are averaged and used to replace the failed zone values. We also impose numerical floors and ceilings on the primitive fields, such that the density, $\rho$, and internal energy density, $e$, are not allowed to drop below 90\% of their initial background values, and the Lorentz factor is not allowed to exceed 20. We also impose relative restrictions between the gas and magnetic field properties, such that $\rho \ge B^2/100$ and $P_\mathrm{gas} \ge P_\mathrm{mag}/25$; these limits help with the stability of the code in the relatively evacuated background region. Whenever mass or energy is added to a cell because of these magnetization limits, this is done in the drift frame according to \citet{Ressler17}.

\subsection{Simulation Setup}

We only simulate the inner region of the Shakura-Sunyaev disk model from $r=4\, GM/c^2$ to $r=160\,GM/c^2$, 0 to $\pi$ in the $\theta$ direction, and from 0 to $\pi/2$ in the $\phi$ direction, making our simulation domain a wedge shape. As these are very thin disks ($H/R \lesssim 0.03$), we use a variety of techniques to concentrate resolution as much as possible toward their inner regions, including a logarithmic radial coordinate, $x_1 = 1 + \ln (r/r_\mathrm{BH})$, a concentrated latitude coordinate,
\begin{equation}
    \theta = x_2 + 0.35 \sin(2x_2) ~,
\end{equation}
and static mesh refinement. We start with a base mesh of $48\times48\times12$ and add two or three levels of refinement focused around the inner disk for an equivalent resolution of $384\times384\times96$ for the highest resolution, four-level simulations. Even so, we only have approximately 6(12) zones per scale height of the disk initially for our three-(four-)level simulations. The three-level grid , used for most of the simulations, is shown in the top panel of Figure \ref{fig:initial_fields}. 

We apply outflow boundary conditions at both radial grid boundaries, reflecting boundary conditions at the poles, and periodic boundaries in the azimuthal direction. One change from how we set the disk up in \citet{Fragile18b} is that here we ignore the relativistic correction terms ${\cal A}$, ${\cal B}$, ${\cal C}$, ${\cal D}$, ${\cal E}$, and ${\cal Q}$ that appear in the \citet{Novikov73} form of the thin-disk solution \citep[see eq. (99) of][]{Abramowicz13}. Otherwise, the procedure is the same.

\subsection{Magnetic Field Setup}

The simulations are each seeded with one of four relatively simple field geometries; the key is that each geometry is qualitatively different in some way. In all four cases, the fields are initially weak relative to the gas and radiation pressure within the body of the disk. Finally, the fields are constructed such that $\beta_{\mathrm{mid,}0}$ is approximately constant with radius. 

\begin{figure}
\centering
\includegraphics[width=0.65\textwidth]{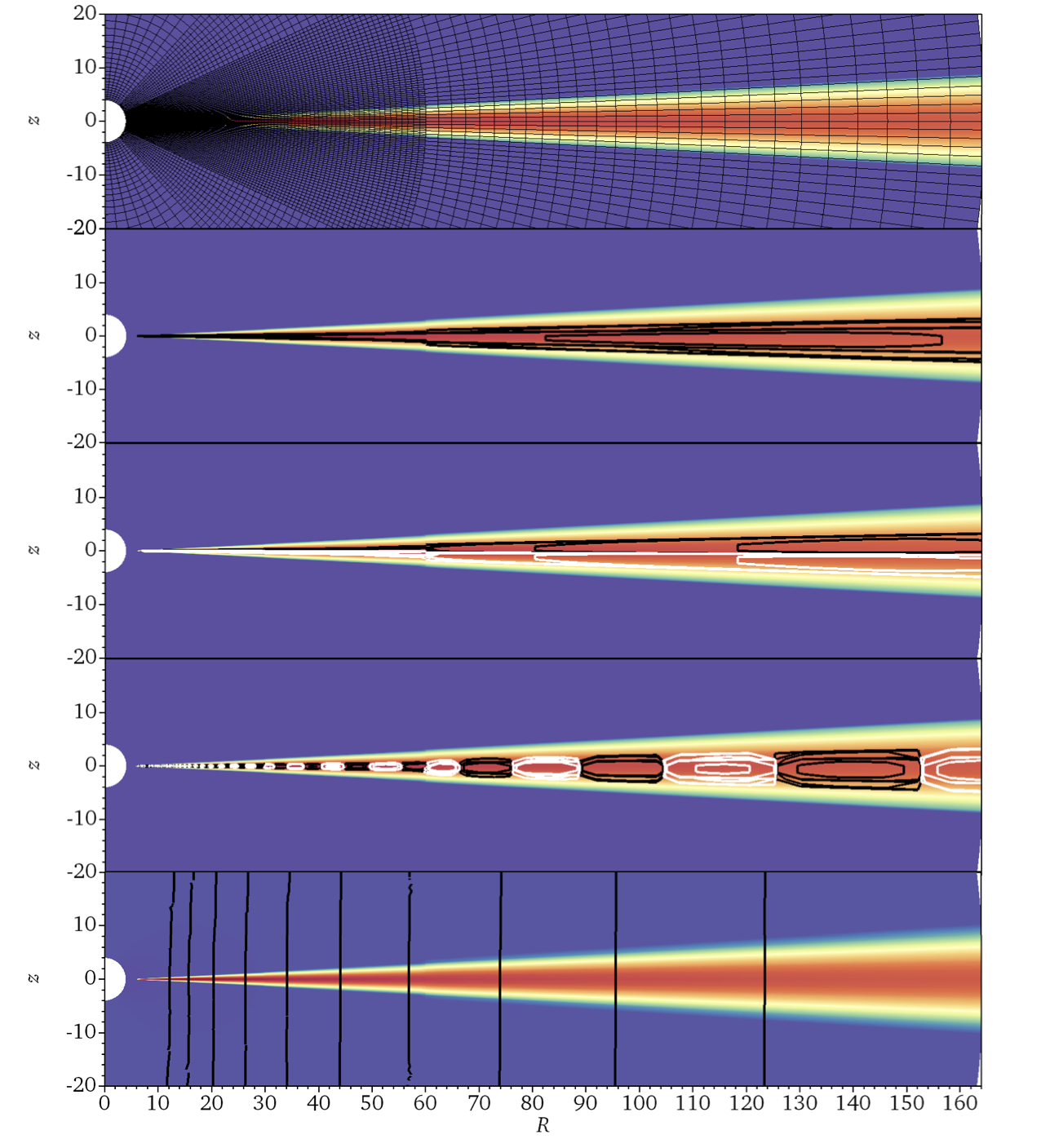}
\caption{Pseudocolor plots of the logarithm of gas density. The {\it top} panel shows the three-level mesh used in most of the simulations. The remaining panels show the different magnetic field configurations that we consider, with black and white lines showing the different magnetic field polarities: zero-net-flux, single-loop or dipole (S3Ed, second panel); zero-net-flux, two-loop or quadrupole (S3Eq, third panel); zero-net-flux, multi-loop (S3Em, fourth panel); and net-flux, vertical-field (S3Ev, bottom panel). We have truncated the vertical extent of each panel to make the details of the disk more easily visible.} 
\label{fig:initial_fields}
\end{figure}

The first geometry we consider is a zero-net-flux, single-poloidal-loop case (second panel of Figure \ref{fig:initial_fields}). This is the standard dipole field configuration that has been used to initialize many global MHD disk simulations; the one difference is that our field is very elongated in the radial direction, extending from near the inner radius of the disk all the way to the outer boundary of our simulation domain. To initialize this field, we first set the azimuthal component of the vector potential to be
\begin{equation}
A_\phi \propto R^{1.5}\frac{\sqrt{e^{(-2.5z^2/H^2)}}\sin\left(\pi R/r_\mathrm{max}\right)}{1+e^\Delta} ~,
\label{eqn:potential}
\end{equation}
where $R=r \sin\theta$ is the cylindrical radius, $H$ is the local height of the disk, $r_\mathrm{max}=30^{1.5}GM/c^2$ is the maximum radius of the grid, and 
\begin{equation}
\Delta=10\left(\frac{z^2}{H^2}+\frac{(R-R_t)^2}{H^2}-1\right) ~,
\label{eqn:delta}
\end{equation}
where $z=r \cos\theta$, $R_t=\mathrm{max}(R_\mathrm{ISCO},R)$, and $R_\mathrm{ISCO}=6GM/c^2$ is the usual ISCO\footnote{Innermost Stable Circular Orbit} radius. We then set the poloidal components of the magnetic field as $\mathcal{B}^r = -\partial_\theta A_\phi$ and $\mathcal{B}^\theta = \partial_r A_\phi$. These choices keep the initial magnetic field confined within our very thin initial disk. This field configuration is subject to a strong radial shear amplification (leading to a growth of the $\mathcal{B}^\phi$ component) due to the orbital motion of the disk (the so-called $\Omega$-dynamo), along with MRI-driven amplification. However, a common feature of all such dipole field configurations is that they have a current sheet exactly at the midplane of this disk. This turns out to be an important factor in determining the subsequent evolution of this case.

Our second magnetic field configuration consists of two poloidal field loops of opposite polarity stacked vertically, one on top of the other, about the midplane of the disk (third panel of Figure \ref{fig:initial_fields}). To achieve this, we use the same vector potential as the dipole field case (Eq. \ref{eqn:potential}), except multiplied by an extra factor of $z$ to introduce the asymmetry across the midplane. Again, we expect significant field amplification from the orbital motion of the disk. Although this configuration introduces a second current sheet, neither is located in the midplane of the disk, unlike the dipole case.

The third field configuration consists of multiple small poloidal loops of alternating polarity distributed in concentric rings moving outward through the disk midplane (fourth panel of Figure \ref{fig:initial_fields}). Each ring has a width comparable to the local disk height. To achieve this, we start from the following vector potential:
\begin{equation} 
A_\phi \propto R^{2}\frac{\sqrt{e^{(-2.5z^2/H^2)}}\sin\left(2\pi R/5H\right)}{1+e^\Delta} ~.
\end{equation}
For such a configuration, we do not expect the field to amplify sufficiently to offer significant pressure support. This is because the narrow radial range of each magnetic cell prevents significant radial shear. Also, this configuration lacks any sort of underlying guide field that can replenish field lost to reconnection. Ultimately, any amplification of this field is limited to the action of the magneto-rotational instability (MRI), which typically saturates at $\beta_r \gtrsim 10$ for zero-net-flux configurations \citep{Turner04,Hirose09}, and therefore, we expect to see thermal runaway (either collapse or expansion) analogous to our earlier simulations with a similar field configuration \citep{Mishra16}. Nevertheless, we run this simulation as a control case. 

For the final configuration, we consider a net-flux, vertical-field threading through the disk (bottom panel of Figure \ref{fig:initial_fields}), such that, again, $\beta$ is reasonably uniform throughout the midplane. Here the vector potential is simply 
\begin{equation}
A_\phi \propto R^{2}.
\end{equation}
Such a field configuration is subject to amplification due to both the shearing at the interface between the disk and background medium and the MRI inside the disk. In non-radiative, shearing box simulations, such a configuration can reach saturation field strengths of $\beta_\mathrm{mid} = P_{\mathrm{gas},0}/P_{\mathrm{mag},0} \approx 0.25$ even for initially weak fields \citep{Bai13,Salvesen16}. If similar saturation strengths could be reached in radiation-pressure-dominated cases, then this may be enough to stabilize the disk \citep{Sadowski16b}.

At the start of all our simulations, the magnetic fields are normalized to match a specific target value for $\beta_{\mathrm{mid}}$. In the dipole, quadrupole, and multi-loop cases, $\beta_{\mathrm{mid}} \approx 100$, while for the vertical field, $\beta_{\mathrm{mid}} \approx 1000$. Since all of our simulations start from the same base disk configuration as the S3E simulation in \citet{Fragile18b}, we follow the same naming convention in this paper. To distinguish the different field configurations, we append the base simulation name with a ``d'' for the dipole case, a ``q'' for the quadrupole case, an ``m'' for the multi-loop case, and a ``v'' for the vertical field case. All of the simulations presented in this work are summarized in Table \ref{tab:params}. The first column shows the model name (S3Ed:dipole, S3Eq:quadrupole, S3Em:multi-loop, S3Ev:net-vertical field, and 4L for four level). The second column shows the duration of each simulation. In third and fourth column we report the initial and late-time ($t=20000\,GM/c^3$) magnetic flux threading the disk. The initial magnetic flux is computed through a sphere of radius $r=15\,GM/c^2$ as
\begin{equation}
\Phi_0 = \int\int{|B^r|r^2 \sin \theta d\theta d\phi} ~.
    \label{eqn:fluxt0}
\end{equation}
Since the dominant field component at late times is the toroidal one, we calculate the late-time flux through a poloidal fan covering $4 \le r/(GM/c^2) \le 15$ and $0 \le \theta \le \pi$ as 
\begin{equation}
\Phi_t = \int\int{|B^\phi| r dr d\theta} ~.
    \label{eqn:flux}
\end{equation}
The flux required to stabilize thin accretion disks such as ours was estimated in \citet{Sadowski16b} to be roughly $2 \times 10^{23}$ G cm$^2$. 
Our initial fluxes are about 4-6 orders of magnitude below this level, while at late times, models S3Eq and S3Ev reach fluxes within 1-2 orders of magnitude of this level. Interestingly, some of our fluxes approach those required for a MAD disk \citep[$\sim 10^{22}$ G cm$^2$;][]{Sadowski16b}, though most of that flux is in the toroidal component. The poloidal flux threading our inner boundary remains well below the MAD limit in all cases. The last column in Table \ref{tab:params} shows the eventual fate of each model.  

\begin{deluxetable}{ccccc}
\tablecaption{Simulation Models \label{tab:params}}
\tablecolumns{5}
\tablehead{
\colhead{Sim} & \colhead{$t_\mathrm{stop}$\tablenotemark{a}} & \colhead{$\Phi_0$\tablenotemark{b}} & \colhead{$\Phi_t$\tablenotemark{c}} & \colhead{Fate} \\
 & \colhead{($GM/c^3$)} & \colhead{($\mathrm{G\,cm^2}$)} & \colhead{($\mathrm{G\,cm^2}$)} & \colhead{}}
\startdata
S3Ed & 30,000 & $1.4\times 10^{18}$ & $1.1\times 10^{20}$ & Unstable\\
 S3Ed4L & 11,076 &  &  &  Unstable\\
S3Eq & 30,000 & $5.6\times 10^{17}$ & $4.8\times 10^{21}$ & Stable(?)\\
 S3Eq4L &  18,645 &  &  & Stable(?)\\
S3Em & 23,988 & $1.0\times 10^{17}$ & $3.6\times 10^{20}$ & Unstable\\
S3Ev &  25,000 & $1.6\times 10^{19}$ & $8.3\times 10^{21}$ & Stable
\enddata
\tablenotetext{a}{Duration of each run.}
\tablenotetext{b}{Initial radial magnetic flux through a shell at $r=15\,GM/c^2$.}
\tablenotetext{c}{Toroidal magnetic flux at $t=20000\,GM/c^3$ through a poloidal fan covering $4 \le r/(GM/c^2) \le 15$ and $0 \le \theta \le \pi$.}
\end{deluxetable}
\section{Stability Results}
\label{sec:results}

In order to fully evaluate the stability of each field configuration, we run each simulation until either the disk clearly collapses (or expands) or until many thermal timescales have passed. In this study, we take the thermal timescale to be $t_\mathrm{th}=0.1(\alpha\Omega)^{-1}$, where $\Omega$ is the local orbital frequency. To standardize our plots, we take $\alpha = 0.02$, which is based on our expectations for the dipole and multipole cases. We expect $\alpha$ to be higher and the true $t_\mathrm{th}$ to be correspondingly shorter in the quadrupole and vertical cases. This estimate of $t_\mathrm{th}$ is a factor of 10 shorter than the estimate we used in \citet{Fragile18b}. The reason is that, without an explicit viscosity in these simulations, there is initially very little heating in the disk. Thus, the thermal energy content is smaller at early times than the simulations in \citet{Fragile18b}. This reduced thermal energy content reduces the time needed for heat to diffuse out (the thermal timescale). For simulations that are able to stabilize themselves, with heating and cooling balancing out and a higher thermal energy content present in the disk, the thermal timescale probably returns to something closer to $(\alpha\Omega)^{-1}$. For all purported stable simulations, we run them for up to $30,000 GM/c^3$, which is over 300 orbits and dozens of (initial) thermal timescales at the ISCO. A big caveat, though, is that this only covers about one viscous timescale, $t_\mathrm{vis} = r^2/\nu = r^2/(\alpha c_s H)$, at that same radius. Another caution is that previous work has shown that the onset of the thermal instability can sometimes be delayed for periods of up to hundreds of orbits \citep{Jiang13,Ross17}, so it is possible that one or more of our stable configurations could turn out to be unstable if they were allowed to run indefinitely.

\subsection{Zero-net-flux, dipole field case (S3Ed)}
\label{sec:dipole}

\begin{figure*}
\includegraphics[width=1\columnwidth]{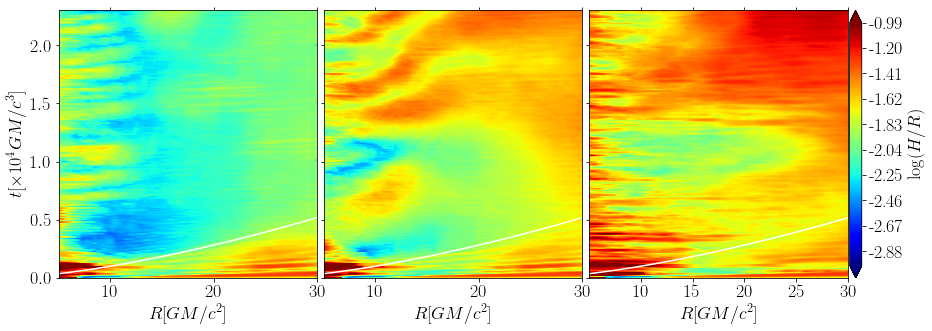}
\caption{Space–time ($R-t$) plots of the density-squared-weighted scale-heights of the disks for the dipole (S3Ed; {\it left}), quadrupole (S3Eq; {\it middle}), and vertical (S3Ev; {\it right}) simulations, showing a vertical collapse for the dipole case and a collapse followed by a recovery for the quadrupole and vertical cases. The solid white curves show the estimated thermal timescale.}
\label{fig:height}
\end{figure*}

It should be of little surprise that all of our simulated disks undergo an initial period of collapse (over roughly a thermal timescale). Remember, they all start off supported by radiation pressure, and since it takes the MRI a few orbital periods to reach saturation, there is a stretch at the beginning of each simulation when there is minimal turbulence, hence little energy dissipation and heating and magnetic pressures are low. As radiation leaks out of the disk and is not replaced during this period (roughly the first thermal timescale), the vertical support declines and the disk height shrinks. In our analysis, the disk height is calculated using a density-squared weighting, as
\begin{equation}
\langle H(R)\rangle_\rho = \sqrt{\frac{\int \rho^2 z^2 \mathrm{d}V}{\int \rho^2 \mathrm{d}V}}~,
\label{eqn:height}
\end{equation}
where the integrals are carried out over each radial shell and $\mathrm{d}V$ is the proper volume of a computational shell. Space-time plots of the scale height (H/R) are provided in Figure \ref{fig:height} for the dipole, quadrupole, and vertical cases.

\begin{figure*}
\centering
\includegraphics[width=\columnwidth]{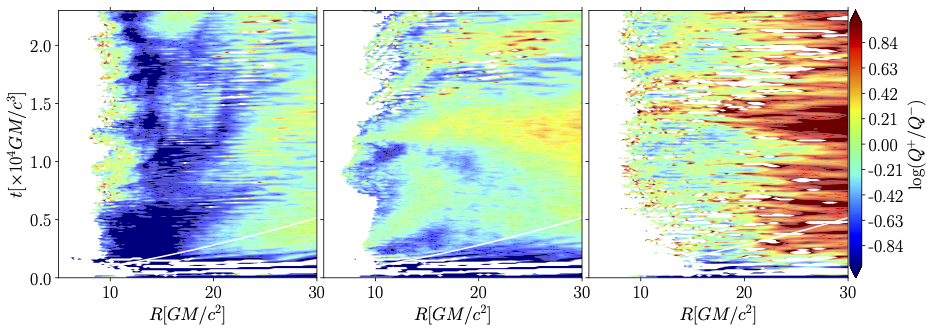}
\caption{Space–time ($R-t$) plots of the ratios of the heating rates, $Q^+$, to the cooling rates, $Q^-$ for the S3Ed ({\it left}), S3Eq ({\it middle}), and S3Ev ({\it right}) simulations, showing that cooling dominates, especially at smaller radii in S3Ed; heating and cooling balance fairly well in S3Eq; and heating seems to mostly dominate, especially in the outer regions in S3Ev. Inside of $\approx 10\,GM/c^2$, the plots are white because the disk is effectively optically thin in those regions, and our heating and cooling formulae do not apply. Beyond that radius, the occasional white patches correspond to locations where the cooling rate takes on negative values, which happens when the flux measured at the photosphere points toward the disk midplane, rather than away. The solid white curves show the estimated thermal timescale.}
\label{fig:heat_cool}
\end{figure*}

The difference between our simulations is the degree to which the disk recovers from this initial collapse. In the case of the dipole configuration, shown in the left panel of Figure \ref{fig:height}, the disk marginally recovers. While the MRI attempts to heat the disk, this heating rate is not quite sufficient to balance the cooling rate, especially at small radii. Figure \ref{fig:heat_cool} shows the ratio of heating, $Q^+$, to cooling, $Q^-$, demonstrating the slight dominance of the latter for the dipole case (left panel). The net heating rate per unit surface area is computed as
\begin{equation}
Q^+(R) = \frac{3}{2}\int\langle V^\phi W_{\hat{r}\hat{\phi}}\rangle_\phi \mathrm{d}\theta ~,
\label{heatingrate}
\end{equation}
where the integration is carried out within the limits of the effective photosphere and the integrand is azimuthally averaged, with $V^\phi \approx \Omega$  the azimuthal component of the fluid three velocity and $W_{\hat{r}\hat{\phi}}$  the covariant $r$-$\phi$ component of the MHD stress tensor in the co-moving frame.  The radiative cooling is computed by tracking the radiative flux through the photosphere at each radius:
\begin{equation}
Q^-(R) = \langle F^\theta_\mathrm{photo+}(R)\rangle_\phi - \langle F^\theta_\mathrm{photo-}(R)\rangle_\phi~,
\label{coolingrate}
\end{equation}
where $F^\theta_{\mathrm{photo}\pm}(R) = -4/3 E_R u_R^\theta (u_R)_t$ is the flux escaping through the top or bottom photosphere. As advective cooling is not important in these simulations, we ignore its contribution to $Q^-$.  Also neglected is the contribution of the radial component of the radiative flux to cooling, which is appreciable only close to the ISCO.

\begin{figure*}
\centering
\includegraphics[width=\columnwidth]{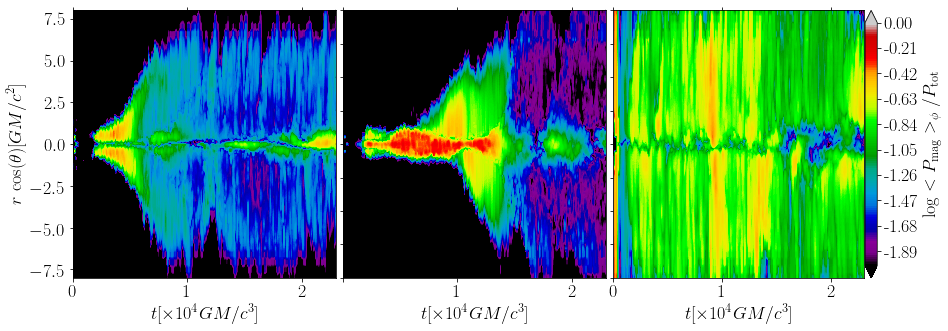}
\caption{Time-space ($t-z$) plots of $\beta_r^{-1}$ (ratio of azimuthally averaged magnetic pressure to total pressure, $P_\mathrm{tot}= \langle P_\mathrm{mag} \rangle_{\phi} + \langle P_\mathrm{gas} \rangle_{\phi} + \langle P_\mathrm{rad} \rangle_{\phi}$) as a function of height in the disks for the S3Ed ({\it left}), S3Eq ({\it middle}), and S3Ev ({\it right}) simulations, evaluated at $r=15\,GM/c^2$. }
\label{fig:beta_z}
\end{figure*}

\begin{figure}
\centering
\includegraphics[width=0.45\textwidth]{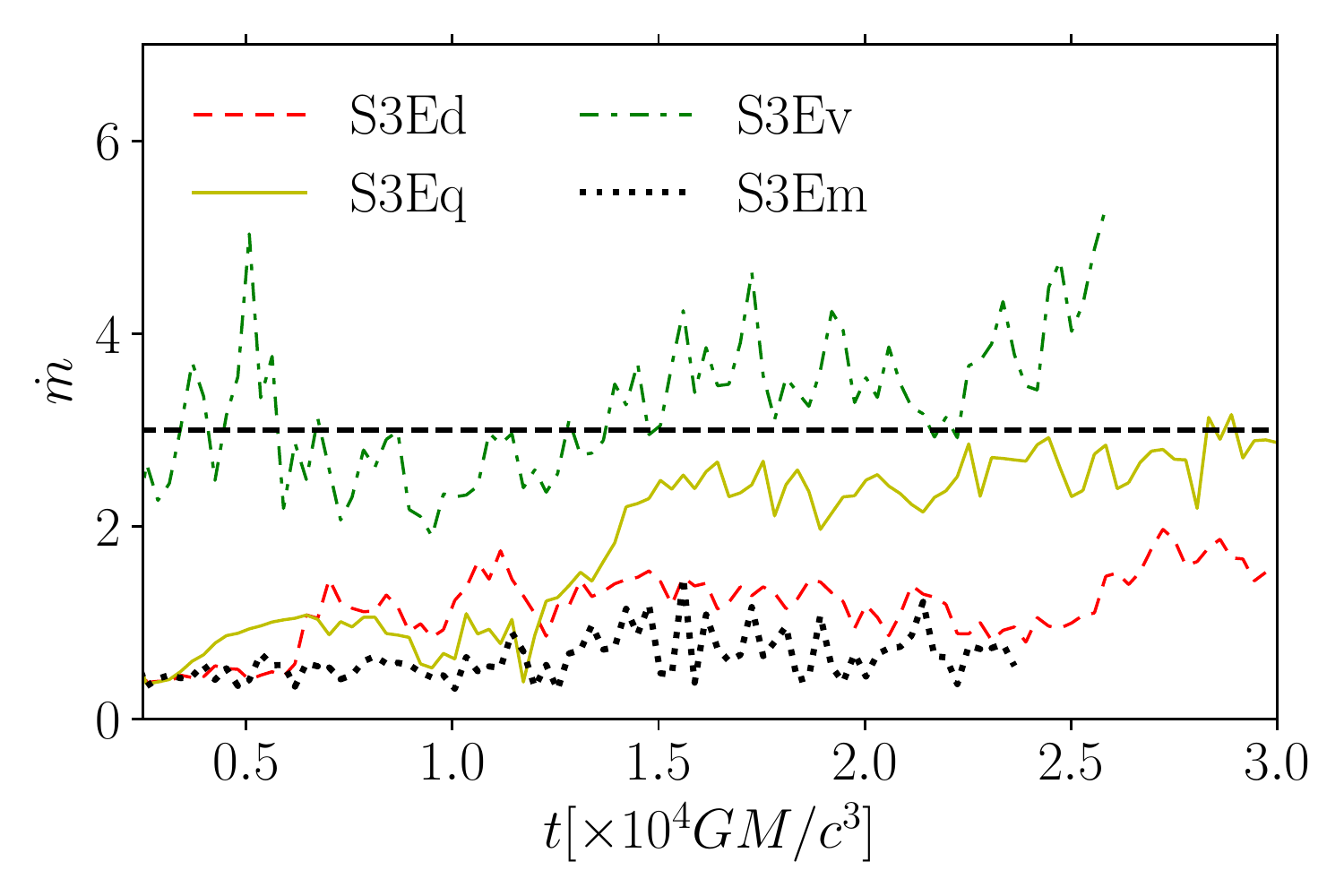} \\
\includegraphics[width=0.45\textwidth]{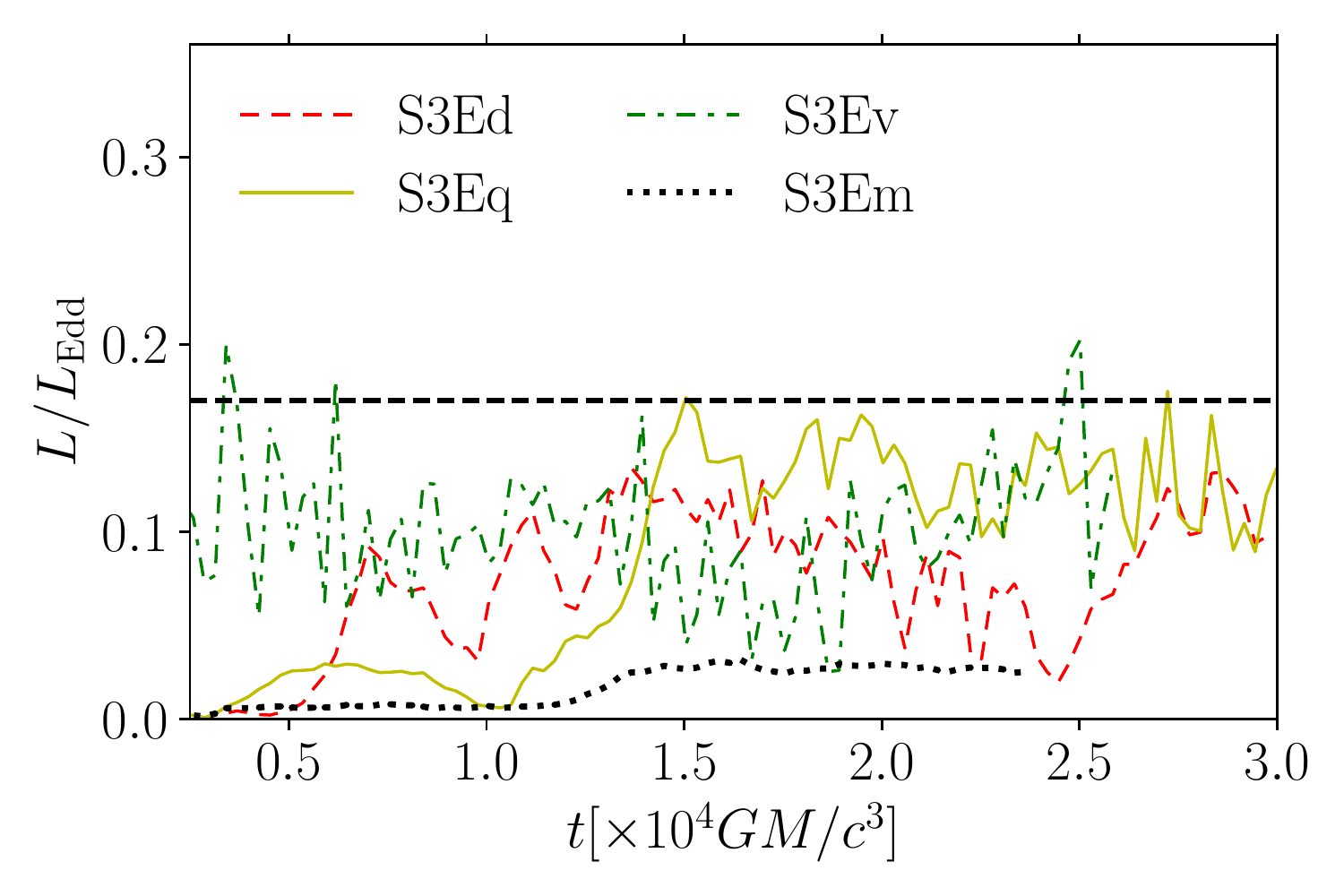}
\includegraphics[width=0.45\textwidth]{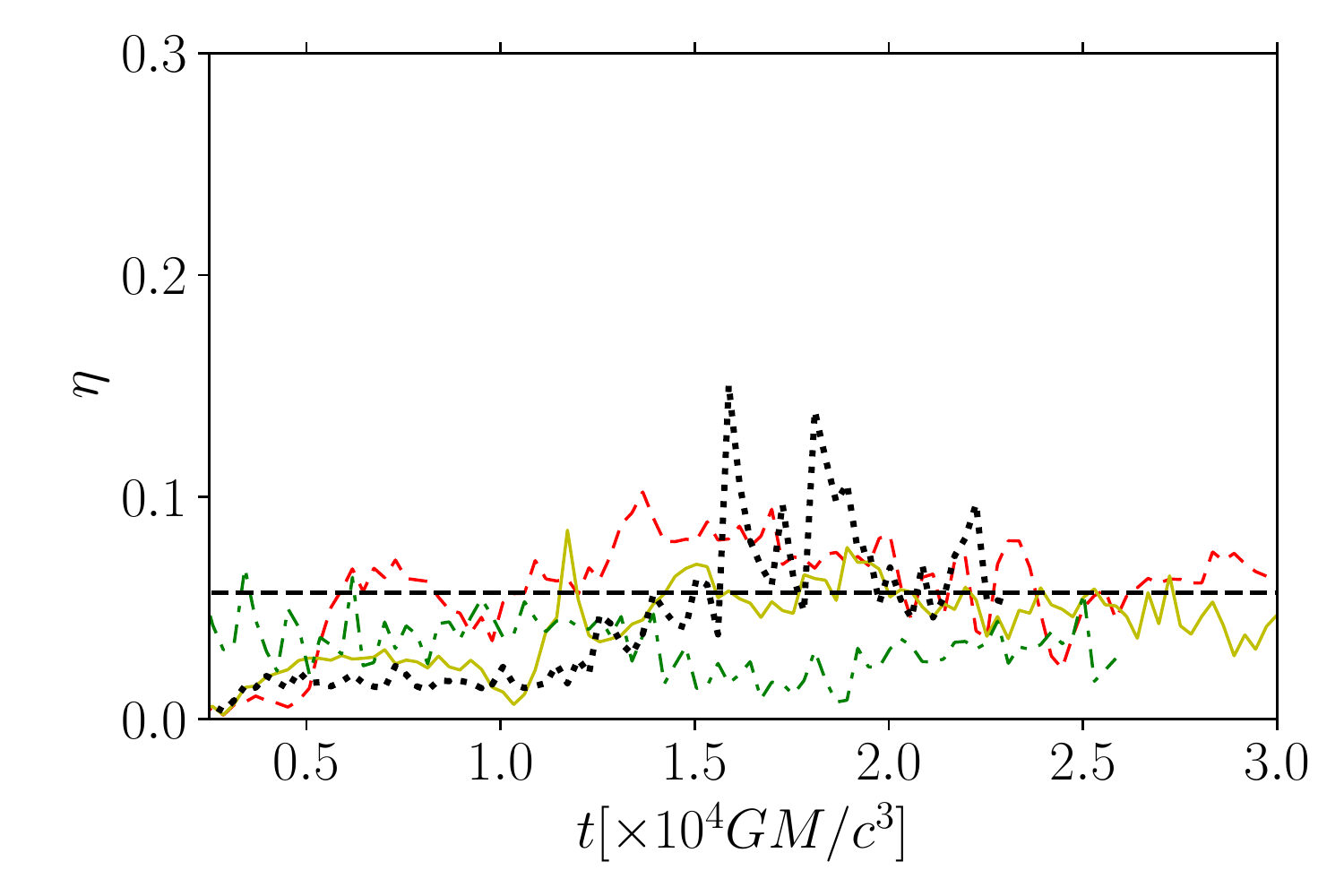}
\includegraphics[width=0.45\textwidth]{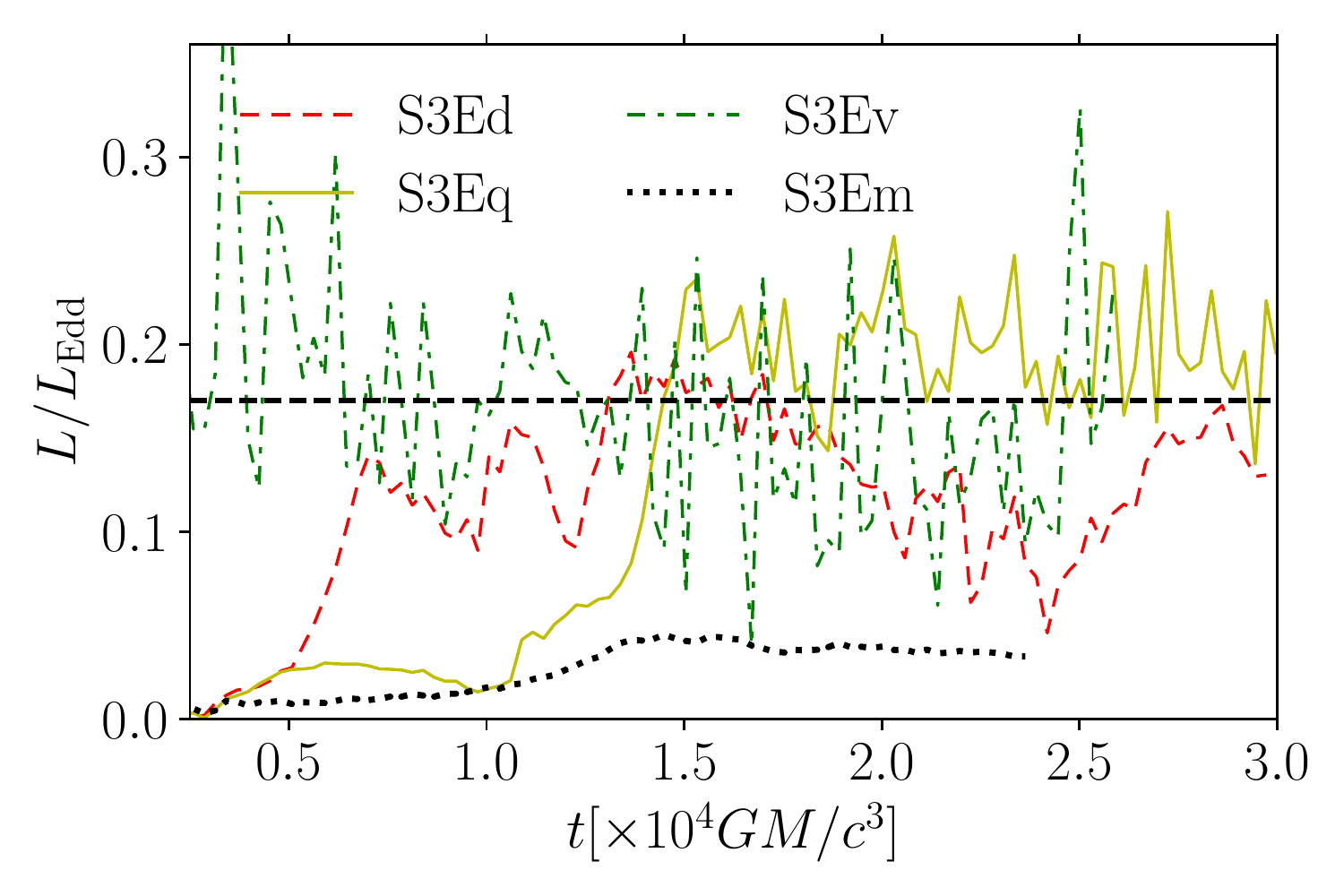}
\includegraphics[width=0.45\textwidth]{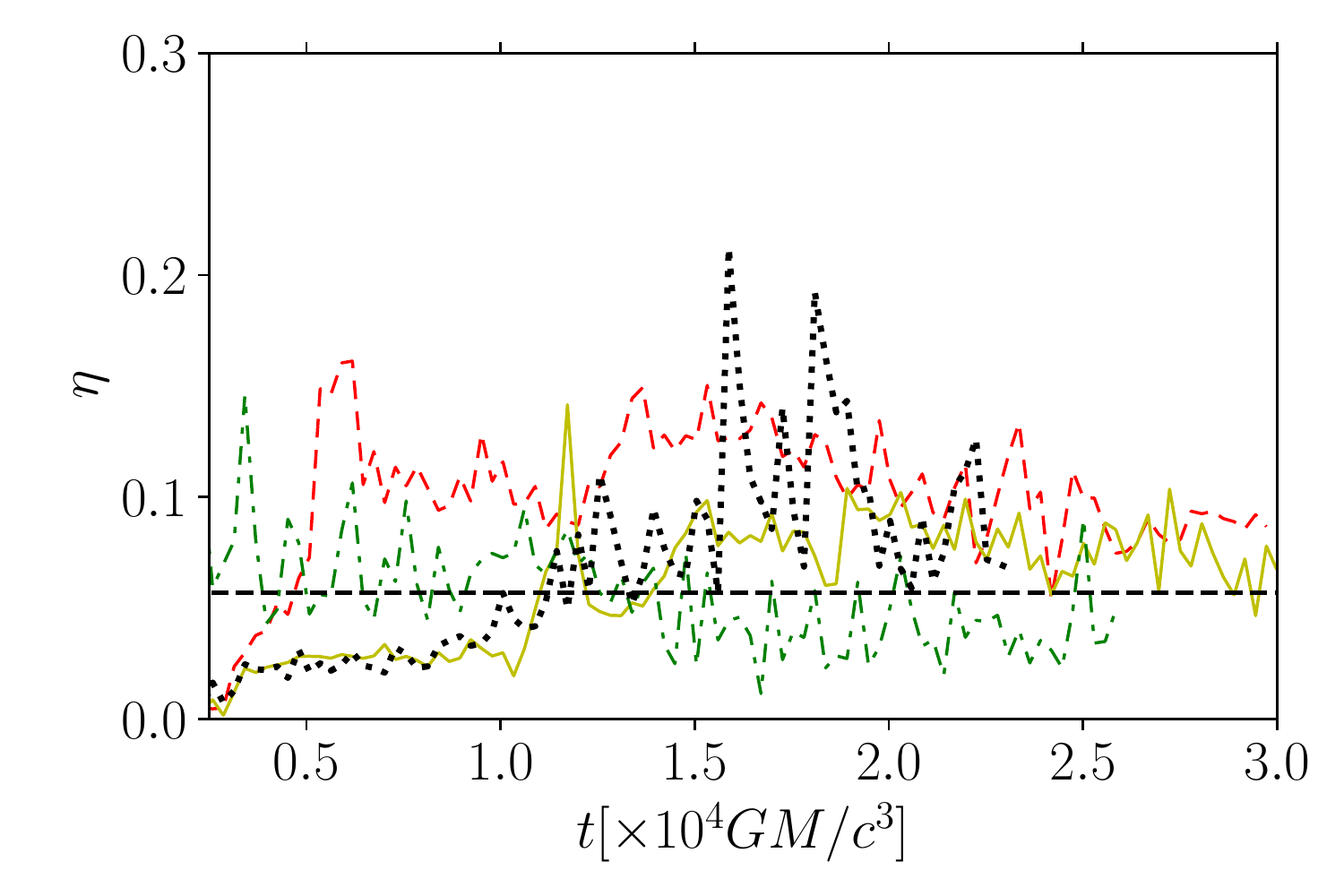}
\caption{Mass accretion rate measured near the ISCO, scaled to the Eddington luminosity, i.e., $\dot{m} = \dot{M}c^2/L_\mathrm{Edd}$ ({\it upper} panel); luminosity through a radial shell at $r \approx 15\,GM/c^2$ ({\it middle left} panel) and $r \approx 20\,GM/c^2$ ({\it lower left} panel), scaled to the Eddington luminosity; and the radiative efficiency, $\eta = L/\dot{M}c^2 = (L/L_\mathrm{Edd})/\dot{m}$ at each radius ({\it rightmost} panels). The horizontal thin, dashed black lines show the target values of $\dot{m}=3$, $L/L_\mathrm{Edd}=0.17$, and $\eta = 0.057$, respectively. Because the unstable simulations (S3Ed and S3Em) have collapsed to a different disk solution, they settle to a lower $\dot{m}$ and $L$. Note that for these plots we employ moving window averages with window widths equal to three consecutive data to smooth them.}
\label{fig:mdot}
\end{figure}

Additionally, the magnetic pressure in S3Ed model fails to reach the stability threshold, $P_\mathrm{mag} > 0.5 P_\mathrm{tot}$, except maybe very close to the disk midplane as shown in Figure \ref{fig:beta_z} (left panel). The S3Ed configuration takes longer to amplify the magnetic field compared to other runs (particularly S3Eq and S3Ev) and ultimately is unable to fully compensate for the lost thermal and radiative support. Instead, the S3Ed disk is effectively ``frozen'' and settles down to a new solution at a lower mass accretion rate and luminosity (Figure \ref{fig:mdot}). However, if the outer parts of the disk are still supplying material at the original, higher rate, as they would be in a real disk or in a much larger and longer simulation, then matter must begin to pool somewhere in the disk. Eventually, this excess material must be accreted, likely in a rapid burst, after which the cycle would likely repeat \citep{John95}. As this type of limit-cycle behavior is not seen in most BHXRBs, this argues against such disks having a predominantly dipole field configuration.

\begin{figure*}
\includegraphics[width=1\columnwidth]{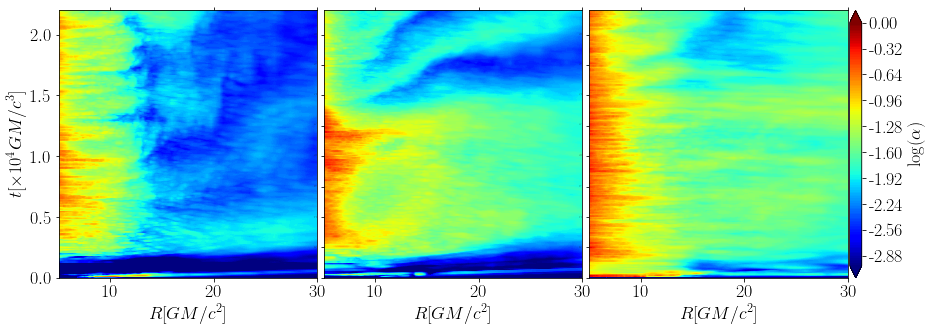}
\caption{Space–time ($R-t$) plots of the density-weighted, shell-averaged viscosity parameter, $\alpha$, for the dipole (S3Ed; {\it left}), quadrupole (S3Eq; {\it middle}), and vertical (S3Ev; {\it right}) simulations.}
\label{fig:alpha}
\end{figure*}

Finally, we find an interesting anti-correlation between the scale height of the disk (Figure \ref{fig:height}) and the effective viscosity, defined here as the density-weighted, height-averaged ratio of the covariant $\hat{r}$-$\hat{\phi}$ component of the stress tensor to the total pressure, i.e., $\alpha\equiv <W_{\hat{r}\hat{\phi}}/P_\mathrm{tot}>_\rho$. A space-time diagram of this quantity is shown in Figure \ref{fig:alpha}. It appears there may even be a threshold value of $\alpha > 0.01$ associated with stability.

\begin{figure*}
\centering
\includegraphics[width=\columnwidth]{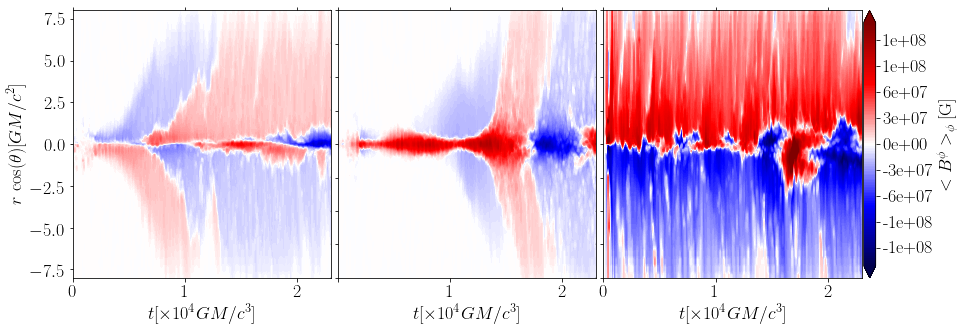}
\caption{Time-space ($t-z$) plots of the azimuthally averaged toroidal component of magnetic field at $r=15\,GM/c^2$. From {\it left} to the {\it right} are models S3Ed, S3Eq, and S3Ev, respectively. Models S3Ed and S3Ev show a current sheet at the midplane, whereas S3Eq does not. Both the S3Ed and S3Eq models show field reversals starting at $t \gtrsim 10^4\, GM/c^3$.}
\label{fig:bphitz}
\end{figure*}

\subsection{Zero-net-flux, quadrupole field case (S3Eq)}
\label{sec:quadrupole}

Unlike the dipole configuration, which seems to quickly collapse to a different configuration, the quadrupole simulation recovers from its initial collapse to re-inflate back to a height comparable to its original profile, as shown in Figure \ref{fig:height} (middle panel). Later ($t > 12,000 GM/c^3$), the disk even ``bounces'' a few times (shown by alternating increases and decreases in height at a given radius).

The notably different behavior of simulation S3Eq compared to S3Ed can best be understood by comparing the left and middle panels of Figure \ref{fig:beta_z}. As mentioned previously, the dipole simulation is plagued by a midplane current sheet at early times that prevents the magnetic pressure from building up sufficiently over the bulk of the disk. The quadrupole simulation, by contrast, does not have a midplane current sheet. In fact, during the initial period of collapse, the midplane value of $\beta_r^{-1}$ actually increases because more magnetic field is being squeezed into a given volume. Without a current sheet to dissipate this energy, the disk is able to store it, and even amplify it, in order to use it later to restore the disk back to something close to its original configuration. More quantitatively, we can see from Figure \ref{fig:beta_z} that simulation S3Eq achieves and sustains the required $\beta_r^{-1} \gtrsim 0.5$ ($-0.3$ on the logarithmic color-bar of Figure \ref{fig:beta_z}), at least until $t \approx 14,000\,GM/c^3$.

However, we see evidence in Figure \ref{fig:bphitz} that, beginning around $6,000 GM/c^3$ for case S3Ed (left panel) and $12,000 GM/c^3$ for case S3Eq (middle panel), the toroidal magnetic field starts buoyantly rising out of the disk. At the same time, the coherent, extended, radial field component, which is crucial for replenishing the toroidal field, begins to break up due to turbulent motions of the MRI. Therefore, the toroidal field is no longer being replaced as fast as it is being lost and $\beta_r^{-1}$ drops. In our higher resolution studies, we found that these same changes occurred, but about 50\% later in time, so the timing of this transition from stability to instability is apparently not fully resolved yet.

Interestingly, there remain periods during the evolution of S3Eq and to some extent S3Ed when a strong radial field component is able to reestablish itself near the midplane, although not always of the same polarity as the initial field. This revived radial field is able to generate sufficiently strong toroidal fields to briefly allow $\beta_r^{-1}$ to again approach the stability limit, but these periods are relatively short and the fields localized; thus, they are not enough to truly restore stability. As a result S3Eq appears to undergo multiple instability cycles, with the disk expanding and contracting vertically on roughly the local thermal timescale. Similar oscillations between stability and instability are suggested in the figures of \citet{Sadowski16}, though the author makes no specific mention of this.

It is interesting that it is only after simulation S3Eq loses its magnetic pressure support and begins to oscillate in height that it reaches a fairly steady state in terms of mass accretion rate and luminosity (Figure \ref{fig:mdot}), with values very close to the targets for this study. During this same period, simulation S3Eq reestablishes a rough thermal equilibrium ($Q^+ \approx Q^-$), as shown in Figure \ref{fig:heat_cool} (middle panel). An initial period of cooling domination ($t \le 2,500\,GM/c^3$) and a later period of heating domination ($10,000 \le t \le 15,000$) are also noticeable. 
We will make a more detailed comparison of this simulation with the Shakura-Suynaev model in Section \ref{sec:Shakura}.

\subsection{Zero-net-flux, multi-loop case (S3Em)}
\label{sec:multiloop}

The zero-net-flux, multi-loop simulation is qualitatively very similar to the simulations we reported in \citet{Mishra16}. It is also the case where the disk is least likely to stabilize, as zero-net-flux MRI turbulence saturates at a field strength of $\beta_r \gtrsim 10$ \citep{Turner04,Hirose09} and there is no global field component for the $\Omega$-dynamo to amplify. Not surprisingly, we see this disk collapse on the local thermal timescale until it becomes too poorly resolved to sustain MRI turbulence. Because of this, we choose not to present any figures specifically for this simulation, although it is included in Figure \ref{fig:mdot}. The solid black curve in Figure \ref{fig:mdot} shows that this model remains under-luminous and also maintains a low mass accretion rate. Our results, plus \citet{Mishra16}, support our conclusion that this field configuration is unstable to thermal collapse in this mass accretion range.

\subsection{Net-flux, vertical field case (S3Ev)}
\label{sec:vertical}

The net-flux, vertical magnetic field configuration leads to our only fully stable disk configuration, in this case by producing the strongest magnetic pressure support among all the models we simulated. It quickly satisfies $\beta^{-1}_\mathrm{r} \gtrsim 0.5$ (right panel of Figure \ref{fig:beta_z}) and has a heating rate that matches, or even exceeds, its cooling rate (right panel of Figure \ref{fig:heat_cool}) and hence causes the disk to elevate (right panel of Figure \ref{fig:height}). 

The strong magnetic pressure support in this case happens despite the presence of a midplane current sheet (seen in the right panel of Figure \ref{fig:bphitz}). This is owing to the very strong magnetic pressure gradients found just above and below the disk midplane, as we will show in the next section. 
\begin{figure*}
\centering
\includegraphics[width=\columnwidth]{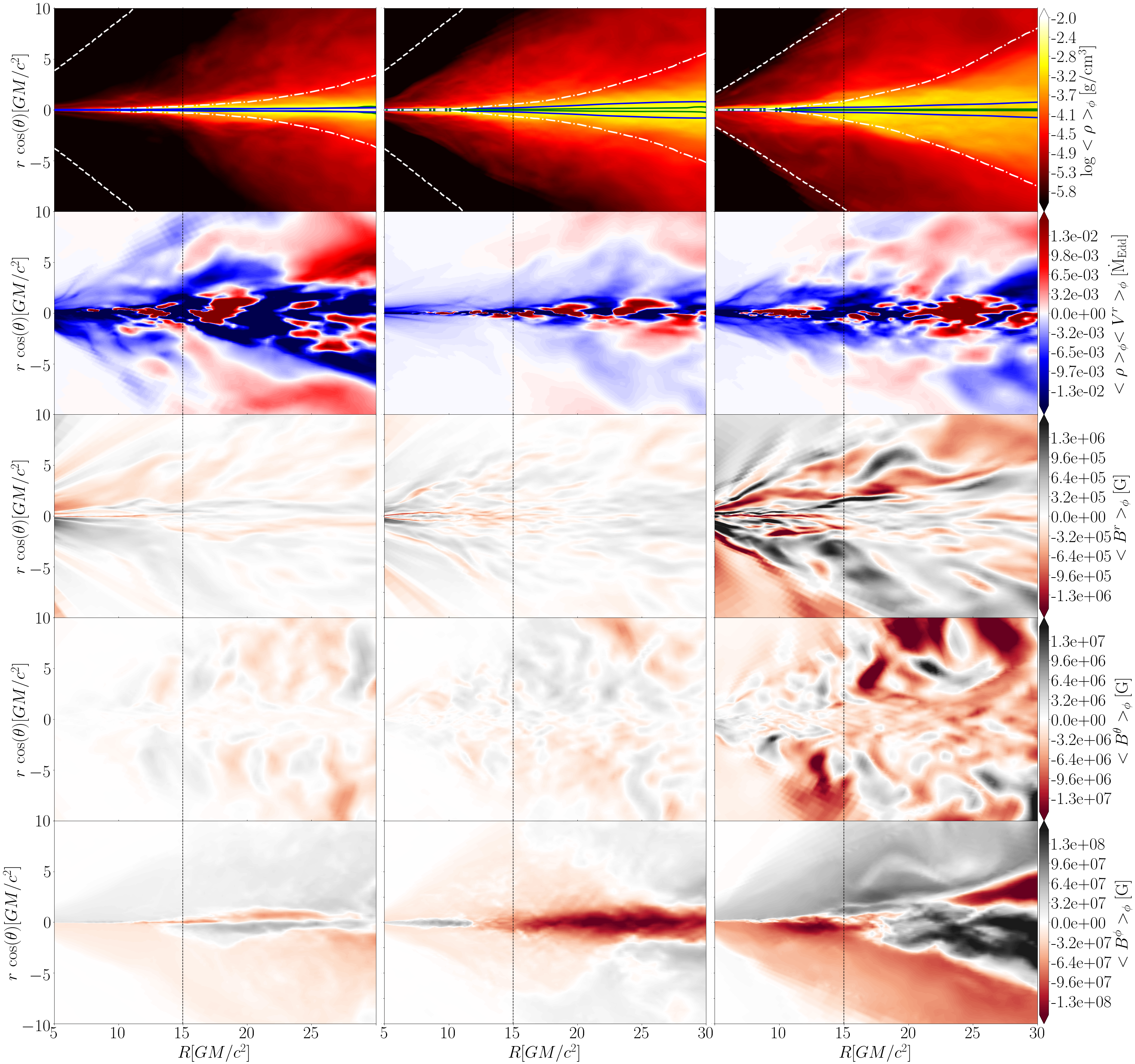}
\caption{Profiles of density ({\it top}), mass flux ({\it second from top}), $B^r$ ({\it middle}), $B^\theta$ ({\it second from bottom}), and $B^\phi$ ({\it bottom}) for the dipole ({\it left}), quadrupole ({\it center}) and vertical ({\it right}) field cases at time $t=20,000\, GM/c^3$. Overlaid on the density profiles are curves representing the time-averaged disk height ({\it solid, blue}), and the absorption ({\it solid, green}), effective ({\it dot-dashed, white}), and scattering ({\it dashed, white}) photospheres. The dashed, black, vertical line corresponds to $R=15\, GM/c^2$ (the radius at which we extract Fig. \ref{fig:beta_z}, \ref{fig:bphitz} and \ref{fig:vertprofileatr}).} 
\label{fig:rzprofile}
\end{figure*}

\begin{figure*}
\centering
\includegraphics[width=\columnwidth]{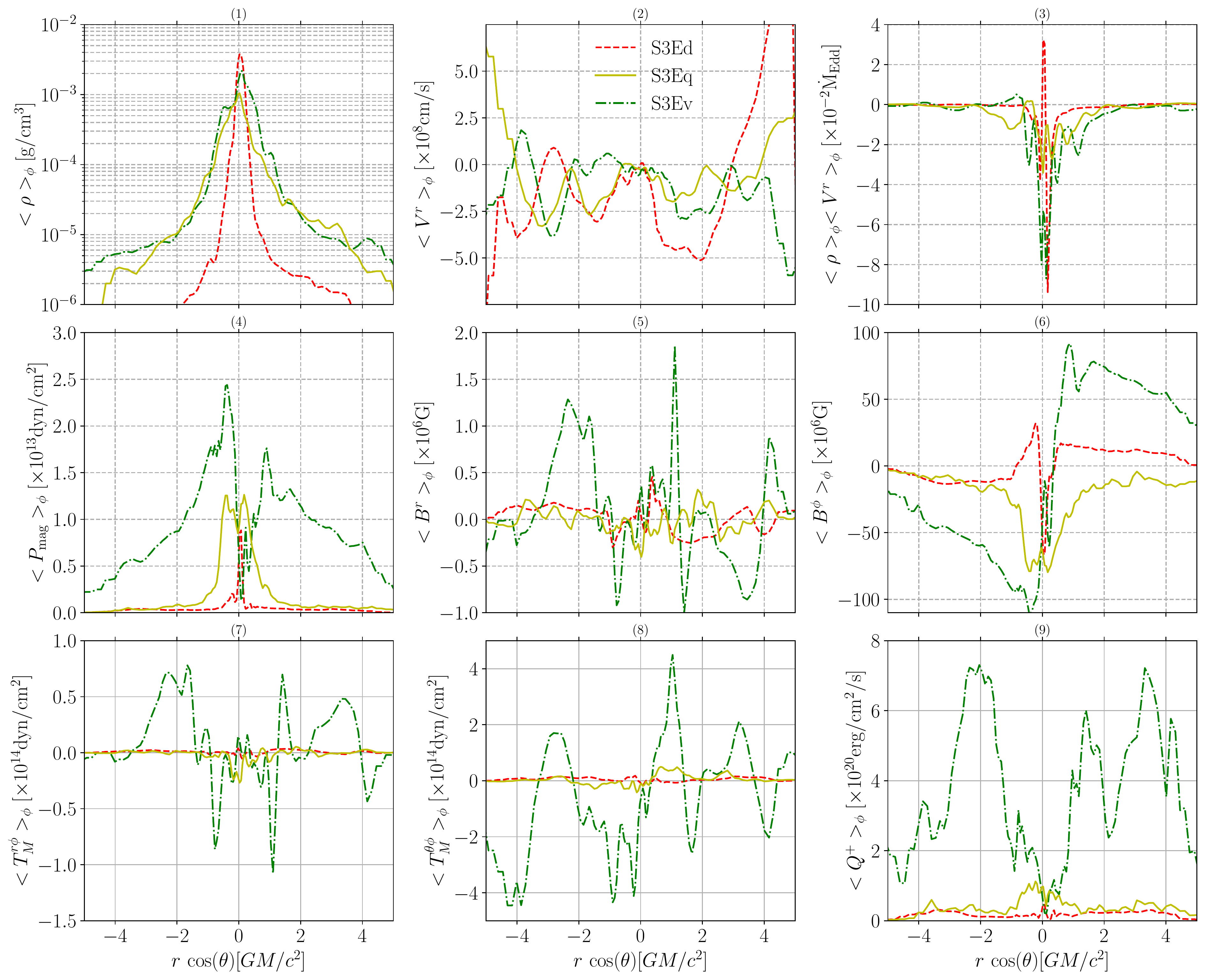}
\caption{Vertical profiles of various disk parameters at $r=15\,GM/c^2$ for the S3Ed ({\it dashed red}), S3Eq ({\it solid yellow}), and S3Ev ({\it dot-dashed green}) simulations at $t=20,000\,GM/c^3$. All quantities are azimuthally averaged.}
\label{fig:vertprofileatr}
\end{figure*}

Another interesting feature of this particular configuration is that it hits the target mass accretion rate and luminosity pretty much right from the beginning (green, dot-dashed curves in Figure \ref{fig:mdot}). This is in contrast to the S3Eq case, which also achieves the target values, but only after $\beta^{-1}_r$ drops below 0.5. This suggests that a net-flux, vertical field configuration does not need to wait for the MRI to develop to begin driving accretion. In fact, the subsequent development of the MRI does not appear to appreciably affect the luminosity. This finding is consistent with previous Newtonian global MHD models reported in \citet{Mishra20}, where the dominant accretion on the surface was driven by the coherent, rather than turbulent, component of the Maxwell stress. 

\section{Disk Profiles}
\label{sec:diskprof}

We now carefully compare the spatial profiles of the dipole (S3Ed), quadrupole (S3Eq), and vertical (S3Ev) field simulations, focusing mostly on the quadrupole and vertical field cases, since those are the ones we found to be stabilized by strong magnetic pressure support. Figure \ref{fig:rzprofile} shows azimuthally averaged profiles of mass density (top panel), mass flux (second panel), and the three magnetic field components (bottom three panels) for each of the three field configurations at $t=20,000\, GM/c^3$. Note that we do not show time-averaged profiles due to the previously mentioned field reversals observed for case S3Eq (middle panel of Figure \ref{fig:bphitz}). Time averaging, especially of the toroidal field, would lead to an incorrect conclusion about its strength.

The density profiles (top panels) for both S3Eq and S3Ev show a puffy structure with a reduced disk midplane density compared to its initial value, a trait that is more obvious at larger radii ($R>15\,GM/c^2$). Although our models cannot scale to a supermassive black hole, such a decrease in the disk midplane density (purely an effect of magnetic pressure increase) could play a role in stabilizing active galactic nuclei disks against gravitational instability \citep{Isaac87,Riols18}.

In each model reported here, the scattering photosphere (dashed, white curves) is quite thick (only visible in the upper and lower left corners of the top panels). The effective photosphere (dot-dashed, white curves) lies inside the disk height (solid, blue curves) at small radii ($R \lesssim 12\,GM/c^2$), but flares out beyond that radius. Since these simulations have not run long enough to reach an equilibrium state much beyond that radius, the flaring could just be a transient feature. Finally, the absorption photosphere (solid, green curves) lies inside the disk height at all radii.

The profiles of mass flux, $\langle \rho \rangle_\phi \langle V^r \rangle_\phi$ (second row), show that, within $R\le15\,GM/c^2$, all of the simulations accrete primarily through the disk midplane, though S3Ed and S3Ev also exhibit significant accretion along the disk surface. The surface accretion is associated with regions of extended radial field coherence seen in the $\langle B^r \rangle_\phi$ plot (third row). For S3Ev, these are regions where the magnetic field doubles back on itself (if you were to follow a single ``vertical'' magnetic field line from the disk midplane to larger heights, you would sees it first bend radially inward and then outward), as noted in previous Newtonian MHD simulations \citep{Zhu18,Mishra20}. Despite such an irregular profile of $\langle B^r \rangle_\phi$, $\langle B^\phi \rangle_\phi$ maintains a strong (two orders of magnitude larger amplitude compared to $\langle B^r \rangle_\phi$), coherent field structure (fifth row, right panel). The S3Eq model also has a strong toroidal field, but with a polarity switch at $R\approx 12\,GM/c^2$. This is interesting because the initial field configuration has its radial component pointing outwards near the midplane, which should lead to a negative component of toroidal magnetic field (as it does for large radii). However, in the inner region ($R\lesssim 12\,GM/c^2$), the disk rearranges the magnetic field leading to a positive toroidal field. This could be better understood by reminding ourselves of the toroidal field reversals seen in the middle panel of Figure \ref{fig:bphitz}. The S3Ev case, on the other hand, has its dominant polarity change roughly in the disk midplane, with stronger $\langle B^\phi \rangle_\phi$ overall compared to model S3Eq. The vertical field component, $\langle B^\theta \rangle_\phi$ (fourth row), has a more complicated, turbulent structure compared to the radial and toroidal field components.  

Continuing our analysis, in Figure \ref{fig:vertprofileatr} we present azimuthally averaged vertical profiles of a number of key disk parameters, all measured at a radius of  $R=15\,GM/c^2$ and at time $t=20,000\,GM/c^3$ for the same three field configurations: dipole, quadrupole, and vertical. In panel (1), the density profile shows that model S3Ed exhibits a much thinner, denser slab profile, whereas the stable configurations, S3Eq and S3Ev, show an enhanced density at higher altitudes, with a corresponding decrease in their midplane values.
In panels (2) and (3), we show radial velocity and mass accretion rate plots. 

The azimuthally averaged magnetic pressure (panel 4) shows one of the key differences between models S3Eq and S3Ev. In the case of S3Eq, the magnetic pressure is highest close to the disk midplane and tapers off from there, whereas the midplane magnetic pressure is relatively low in the case of S3Ev because of the current sheet there. But on either side of the midplane, the magnetic pressure is quite high. In fact, it is perhaps a bit surprising that such different magnetic pressure profiles could yield relatively similar density profiles (panel 1). This is explained by the fact that the radiation pressure is still a major contributor, and has a roughly similar profile in both cases (not plotted).

In panels (5) and (6) we show the radial and azimuthal magnetic field components. The radial magnetic field shows a complicated structure with multiple reversals in the S3Eq and S3Ev cases. The azimuthal magnetic field is two orders of magnitude larger in amplitude and maintains a coherent structure showing no field reversals over this height in the S3Eq case and only one field reversal at the disk midplane in the S3Ev case.

In panel (7), we show $\langle T^{\mathrm{r}\phi}_M\rangle_\phi = -\langle B^r\rangle_\phi \langle B^\phi\rangle_\phi$, which is the coherent component of the Maxwell stress tensor. Near the disk midplane, all the models show very little coherent field. This suggests that the disk region is turbulent, which hinders the development of coherence. The S3Ev model, though, develops coherent magnetic field at higher altitudes, though the sign of the stress reverses multiple times. No such behavior is seen in the other two models. In panel (8) we show the vertical component of the coherent stress, which is again very small for cases S3Ed and S3Eq, whereas S3Ev again exhibits large fluctuations at higher altitudes.

In panel(9), we show the vertical profile of the heating rate. Interestingly, the S3Eq and S3Ev cases have nearly identical heating rates in the disk midplane. At higher altitudes (around $z \approx \pm 2.5\,GM/c^2$), the S3Ev case has about an order of magnitude larger heating compared to S3Eq and S3Ed. These results inform the longstanding question of where the maximum dissipation occurs within a disk. Standard accretion disk theory assumes that disk heating is primarily confined to the disk midplane, as exhibited by our S3Eq case. This is consistent with its magnetic pressure profile. Contrarily, the S3Ev model shows the least heating in the disk midplane (yet still equal to the S3Eq case), but greatly enhanced heating in the region $1 \lesssim \vert z \vert/(GM/c^2) \lesssim 4$. If we compare panels (6) and (9), we see that the enhanced heating rate correlates roughly with the amplitude of the toroidal magnetic field. In the S3Eq case, the strongest toroidal magnetic field is confined within the disk region with a maximum at the disk midplane, whereas in the S3Ev case, the toroidal magnetic field has a local minimum at the disk midplane (due to the current sheet there), while it is larger at higher altitudes. Although, these are ideal GRMHD models, the heating due to magnetic energy dissipation still scales with the available magnetic energy in a given region. The enhanced magnetic energy in the S3Ev case means there is more available energy to cascade down; hence, the enhanced heating in this case. This enhanced heating may be compensated by extra cooling due to outflows (some of which are seen in the second row of Figure \ref{fig:rzprofile}) that help maintain thermal stability \citep{Li14}.

\section{Comparing Our Stable Solutions to Shakura-Suynaev}
\label{sec:Shakura}

We find that, despite their different magnetic field setups and evolution, both the S3Eq and S3Ev configurations can lead to stable disks supported primarily by magnetic pressure, though the S3Eq model seems to fluctuate between stability and instability. Both cases achieve mass accretion rates and luminosities close to what would be expected based on our starting Shakura-Sunyaev model. 
We now compare how well our simulated disks match other predicted properties of the Shakura-Sunyaev model. Since the Shakura-Sunyaev model is a one-dimensional, vertically integrated model without explicit turbulence, we consider time-averaged, radial profiles of our simulations to be the closest proxy. Three key properties to focus on are the disk height, temperature, and surface density. 

\begin{figure}
\centering
\includegraphics[width=0.4\textwidth]{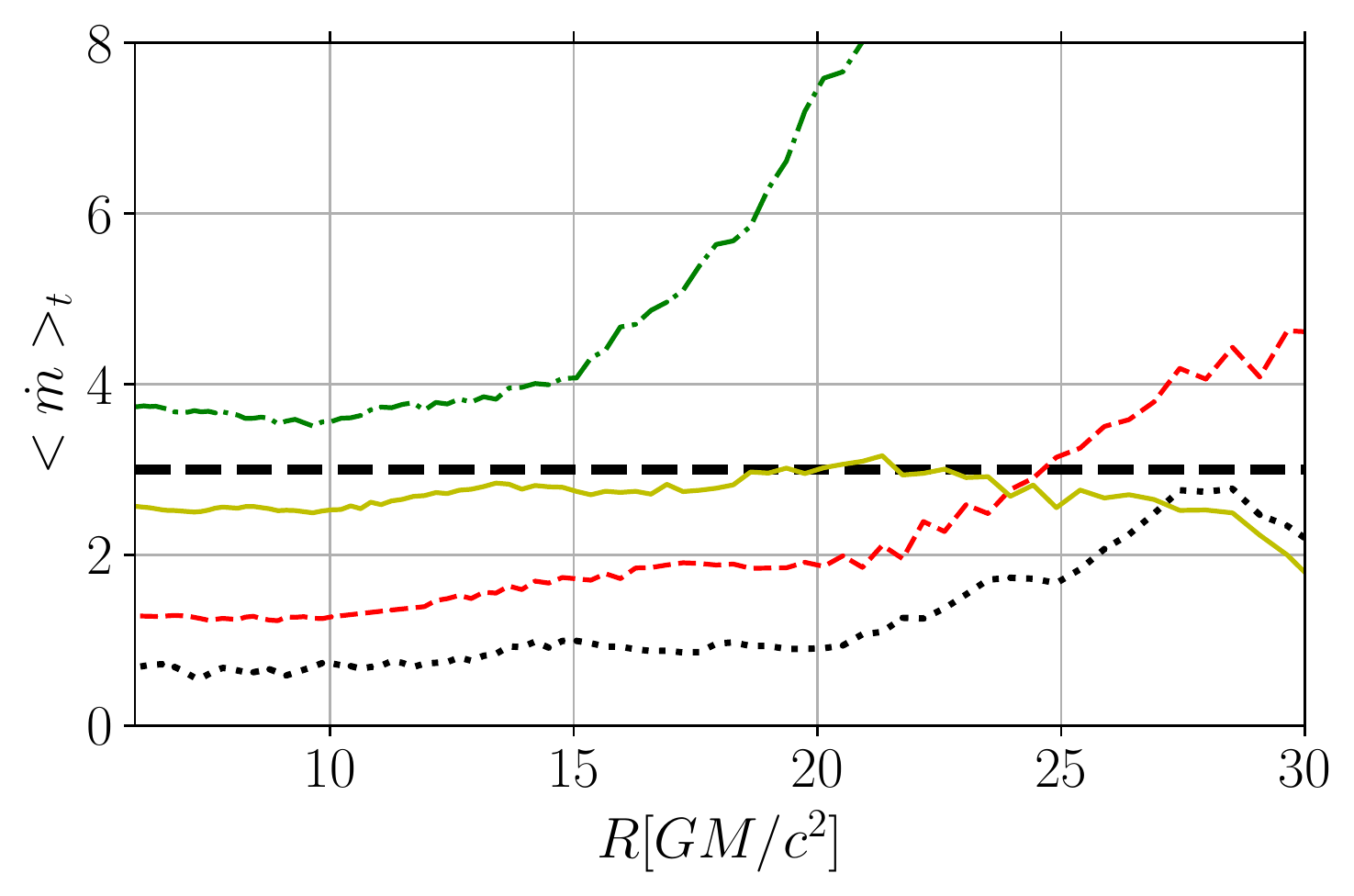}
\includegraphics[width=0.4\textwidth]{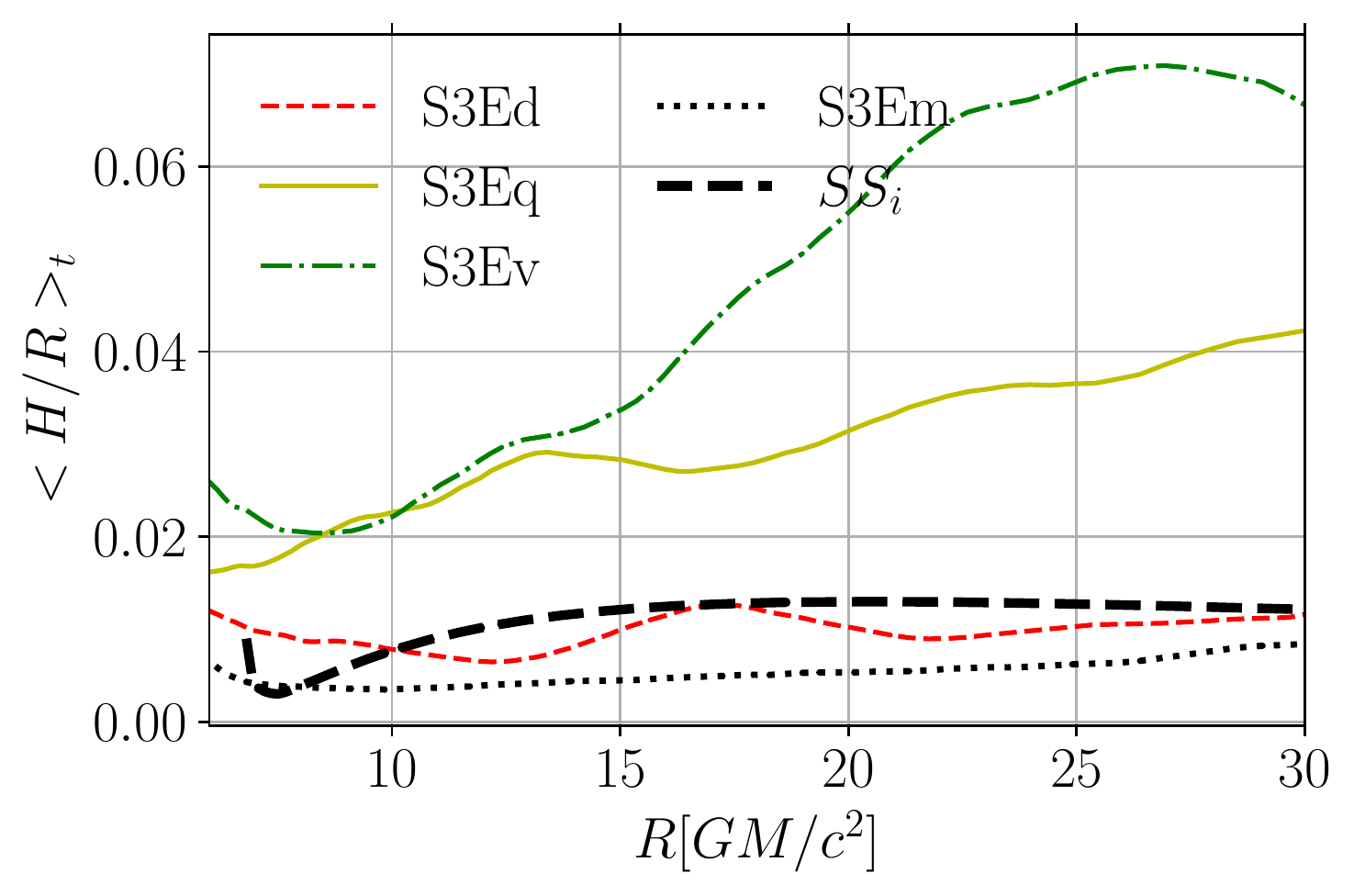}
\includegraphics[width=0.4\textwidth]{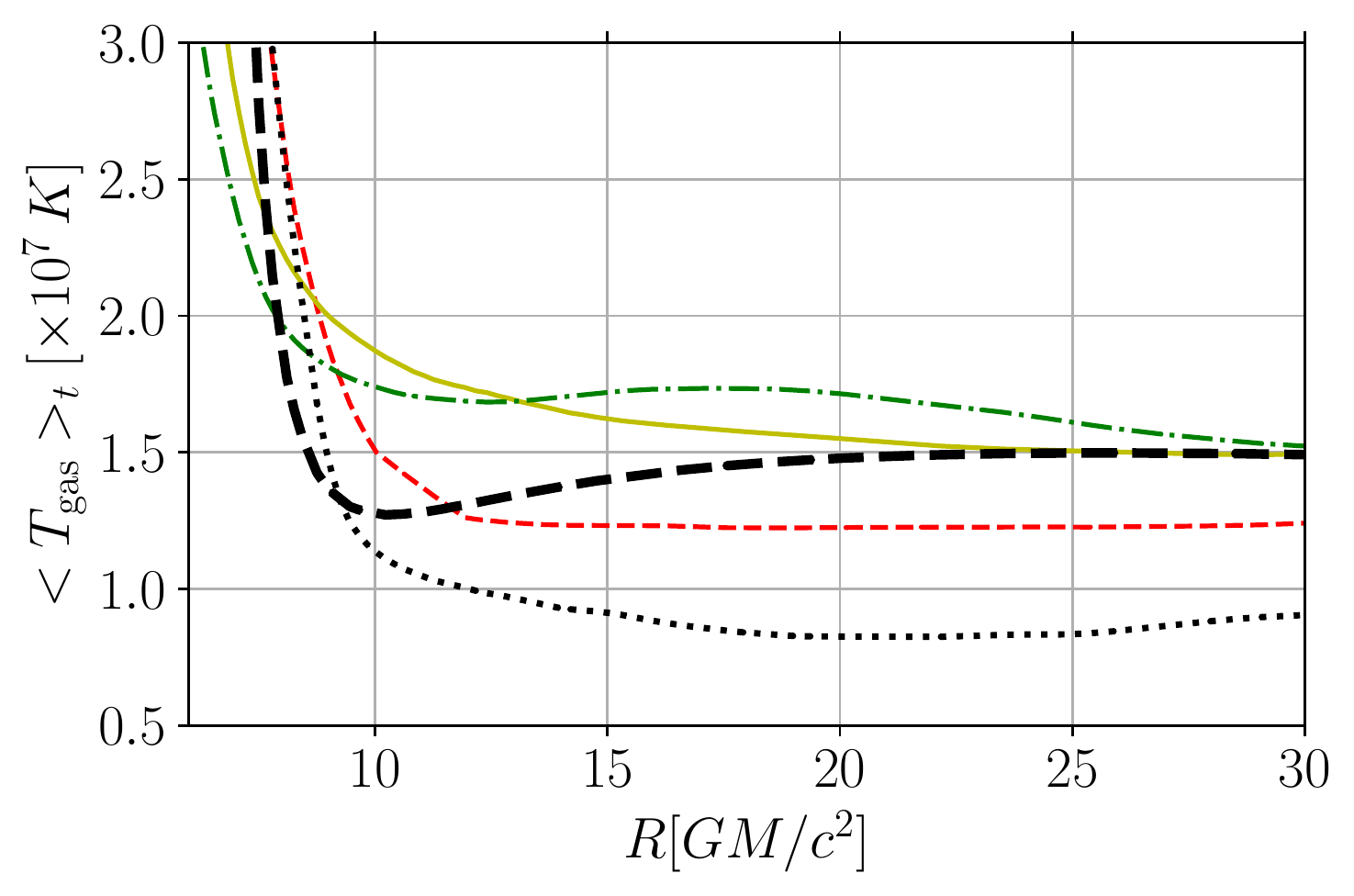}
\includegraphics[width=0.4\textwidth]{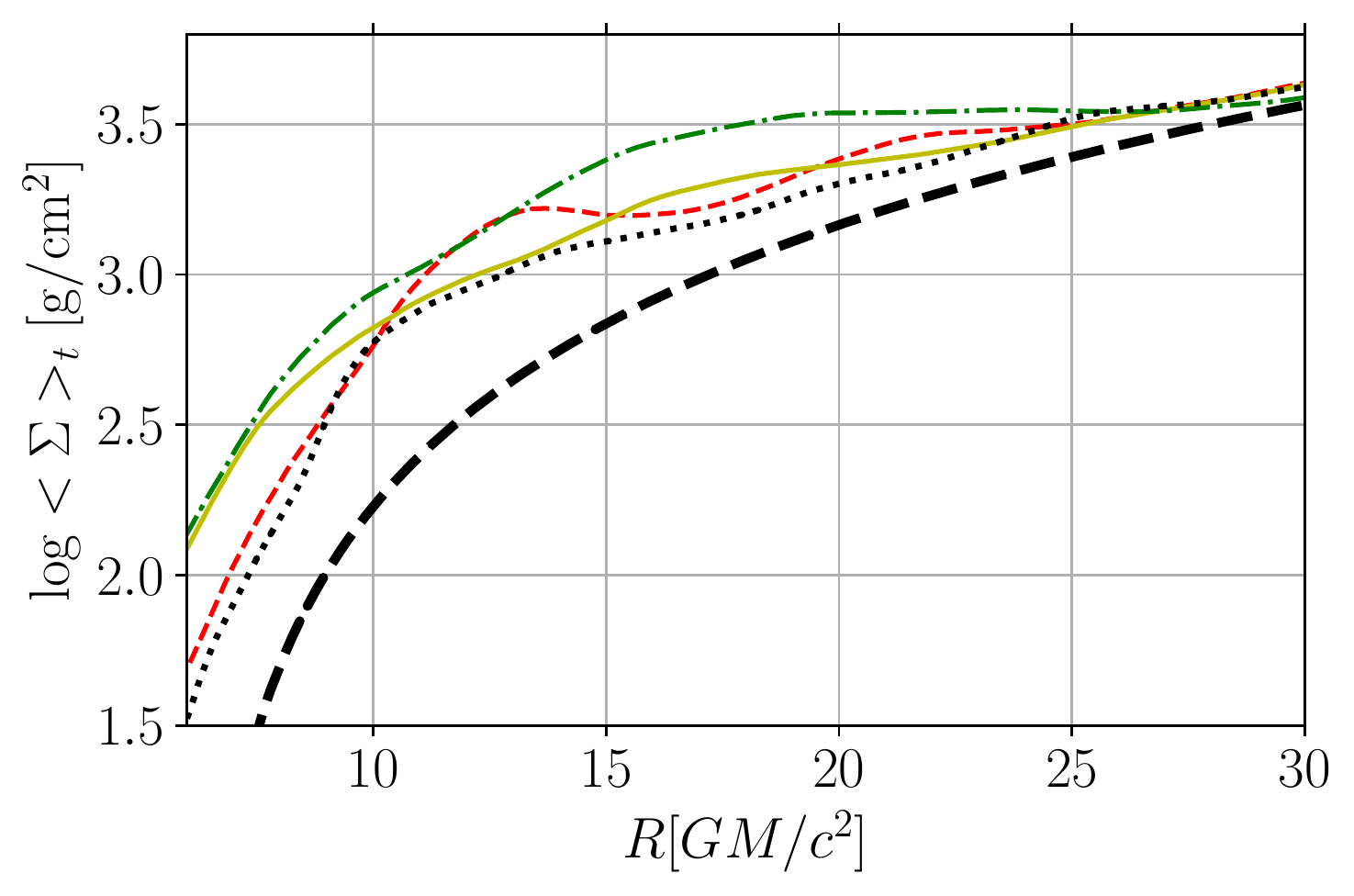}
\caption{Comparisons of the Shakura-Sunyaev disk model ({\it black dashed lines}) with profiles from the simulations for the mass accretion rate ({\it upper left}), disk height ({\it upper right}), gas temperature ({\it lower left}), and surface density ({\it lower right}), time-averaged over the period $t=18,000 GM/c^3$ to the final time of each simulation as reported in Table \ref{tab:params}. Our stable disk models, S3Eq and S3Ev, show a height profile somewhat thicker than the model prediction. The temperature remains nearly the same as what we started with, while the surface density finishes slightly higher than its initial configuration. Note that the dashed black curves are computed using the initial disk profiles from the numerical model.}
\label{fig:SS8}
\end{figure}

In Figure \ref{fig:SS8}, we provide azimuthally and time-averaged radial profiles of the mass accretion rate, $\dot{m}$; disk height, $H$, defined as in eq. (\ref{eqn:height}); disk gas temperature, $T_\mathrm{gas}$, defined as 
\begin{equation}
\langle T_\mathrm{gas}(R)\rangle_\rho = \frac{\int \rho T \mathrm{d}V}{\int \rho \mathrm{d}V}~;
\label{eqn:tgas}
\end{equation}
and disk surface density,
\begin{equation}
\Sigma = \int \frac{\langle \rho \rangle_\phi d\theta}{r dr}~.
\label{eqn:sigma}
\end{equation}
From the plot of $\dot{m}$, we can see that each simulation has achieved a steady state out to about $R\approx 15\,GM/c^2$, with S3Eq and S3Ev closely straddling the target value. 
The S3Ed and S3Em models show lower accretion rates, consistent with their collapse. The upper right panel compares the disk scale height with the standard disk model (black dashed curve). We can see that both of the stable disk configurations (S3Eq and S3Ev) achieve heights much larger than the standard disk model, with a disk scale height of $H/R \approx 0.03$ at $R=15\,GM/c^2$, which is nearly a factor of two larger than predicted by the Shakura-Sunyaev model (although the higher resolution, 4-level model, S3Eq4L, is somewhat thinner). This thickened structure is in agreement with previous GRRMHD simulations of black hole accretion disks at comparable accretion rates \citep[e.g.][]{Sadowski16,Debora19,Maciek22}. S3Ed (and even more so S3Em), on the other hand, has collapsed below the expected disk scale height. The lower left and right panels show profiles of $T_\mathrm{gas}$ and $\Sigma$. We notice that all our disk configurations have temperature profiles nearly the same as the Shakura-Sunyaev model. The profiles of $\Sigma$, by contrast, show elevated surface densities (factors of 3-10) compared to the standard model.

\section{Comparison to Previous Simulations}
\label{sec:Sadowski}
We already mentioned that our zero-net-flux, multi-loop simulation, S3Em, is very similar to the simulations we presented in \citet{Mishra16}. The biggest difference is in the initial setup of the disk. In \citet{Mishra16}, we initialized the disk as a constant height slab of gas, orbiting everywhere at the Keplerian frequency. As such, the simulation was in some ways more akin to a radially extended shearing box simulation than a traditional global one. A negative consequence of that choice was that the disk did not start on the thermal equilibrium curve for a standard disk. This made it more difficult to assess the true nature of the instabilities we witnessed, though a thermal collapse was clearly evident. This is fixed in the current paper by starting from the Shakura-Sunyaev solution, which by construction, begins on the thermal equilibrium curve. Despite the differences, both sets of simulations reach the same conclusion: that a zero-net-flux, multi-loop magnetic field configuration is unable to stabilize itself in the radiation-pressure-dominated regime.

A couple of other simulations we performed share some basic properties with simulations presented in \citet{Sadowski16}. Specifically, our zero-net-flux, dipole case, S3Ed, is very similar to simulation ``D'' of that paper, and our zero-net-flux, quadrupole case, S3Eq, is very similar to their simulation ``Q.'' The biggest difference is that \citet{Sadowski16} started each simulation from a torus of gas initially located far ($R \gtrsim 40\,GM/c^2$) from the black hole. With the build up of MRI turbulence, that torus spreads out into a flatter, wider disk. Although this is a popular starting configuration, it does have its shortcomings when it comes to the goals of studying thermal stability. One is that it is nearly impossible, {\em a priori}, to know what mass accretion rate one will get from such simulations; it is a matter of trial and error. Another is that the mass accretion rate usually decays over longer timescales, as the torus eventually depletes of mass. Finally, this configuration can only achieve inflow equilibrium inside of roughly the initial mid-radius of the torus. To avoid these issues, and to more directly compare with theoretical predictions, we chose instead to initialize our simulations from the Shakura-Sunyaev disk solution \citep{Shakura73}, with weak magnetic fields added. This gets us around the shortcomings of the torus configuration in that we set the mass accretion rate as one of our input parameters, matter can be continually fed in from the outer boundary of the simulation domain, and an inflow equilibrium can, in principle, be reached over most of the grid.

It is noteworthy then, that despite these differences in the setup, we reach essentially the same conclusions as \citet{Sadowski16}, namely that the dipole configuration is unstable, while the quadrupole one is marginally stable. \citet{Sadowski16} concluded that the dipole configuration was unstable based upon: 1) a very thin profile (low $H$) of the disk; 2) a dominance of cooling over heating; 3) a lack of magnetic pressure support; and 4) a drop in $\dot{m}$ -- the same points we made in Section \ref{sec:dipole}. Conversely, \citet{Sadowski16} concluded that the quadrupole configuration was stable based upon: 1) a balance of heating and cooling; 2) magnetic pressure domination, $\beta_r^{-1} \gtrsim 0.5$; and 3) a steady $\dot{m}$ -- the same points we made in Section \ref{sec:quadrupole}.

These same two simulations, S3Ed and S3Eq, also bear some similarities to the two simulations presented in \citet{Jiang19}, in that one of the \citet{Jiang19} simulations, AGN0.2, started with a dipole field configuration, while the other, AGN0.07, used a quadrupole. Both were initialized from a torus configuration similar to \citet{Sadowski16}, although the \citet{Jiang19} simulations were tuned to active galactic nuclei (AGN) parameters (e.g., $M=5\times10^8 M_\odot$). Nevertheless, they cover similar ranges of mass accretion rate and luminosity as our simulations, and are, thus, relevant to a discussion of thin-disk stability. Although \citet{Jiang19} claim that both of their simulations are thermally stable, there are hints in Figures 1 and 2 that their AGN0.2 (dipole) model undergoes a transition around $40,000 GM/c^2$, where $\dot{M}$ and $L$ increase (by a factor of 2 in the case of $L$), the midplane density jumps up (by an order of magnitude), and the gas temperature inside the disk drops (also by about an order of magnitude). All of this happens around the same time $B^\phi$ undergoes a major field reversal. Most of this is reminiscent of the transition we see in our S3Eq simulation around $12,000 GM/c^3$, which we associate with a transition from stability to instability. Interestingly, the quadrupole simulation in \citet{Jiang19} appears to remain stable until at least $45,000 GM/c^3$, nearly four times longer than our S3Eq simulation and about 2.5 times longer than S3Eq4L. Since the \citet{Jiang19} simulations were done at a higher effective resolution than ours, this could just be an extension of the effect we noticed that higher resolution simulations remain stable longer. Or it could be a product of other differences, such as how the radiation is handled in each case.

In addition to S3Ed and S3Eq, we also modeled an initial magnetic field configuration with net-vertical magnetic flux (S3Ev). Such net-flux configurations have been gaining interest in recent years because of their producing higher effective viscosities \citep{Hawley95,Bai13,Salvesen16} and feeding dynamically important magnetic fields toward the central object \citep{Igumenshchev08,Beckwith09,Cao11}. Due to numerical challenges in simulating such a magnetic field configuration, there are very few reported global simulations of them, especially involving thin disks; a couple examples are \citet{Zhu18} and \citet{Mishra20}. Our S3Ev model shows rapid magnetic field amplification and surface accretion with weak outflows along the disk midplane, similar to those reported in \citet{Mishra20}. 


\section{Discussion and Conclusions}
\label{sec:conclusion}

We performed four simulations of Shakura-Sunyaev thin disks threaded with different magnetic field configurations to evaluate their thermal stability and confront these models with observations of stable disks. Our zero-net-flux quadrupole and net-flux vertical field configurations seem to have achieved the desired stability (at least for some period). Both evolve to a late time accretion rate of $\approx 3 L_\mathrm{Edd}/c^2$ and luminosity of $\approx 0.17 L_\mathrm{Edd}$. The properties of these simulated disks are broadly consistent with the Shakura-Sunyaev model, with turbulent heating largely matching radiative cooling inside a magnetic-pressure-supported thin disk. As mentioned previously, it could be that the thermal instability would manifest itself on longer timescales (hundreds of orbital periods and many thermal timescales). However, we have confirmed that our stable configurations can reach $\beta_r^{-1} \gtrsim 0.5$, which should remove the instability.

One limitation of our simulations is that they only cover a quarter of the azimuthal domain. This prevents us from capturing low-order azimuthal structures such as the spiral features seen in \citet{Mishra20}. These could substantially alter the density profile of the disk and hence their radiative properties.

Another issue is that all of our simulations undergo an initial thermal collapse before some of them recover. In order to prevent this initial transient feature, we could have started our simulations from an already turbulent disk to overcome the initial imbalance between heating and cooling. However, it is unlikely in our opinion that this would have changed the outcomes of any of the models. The key to stabilizing these disks is their ability to rapidly build up and sustain a large magnetic pressure. Simulations without a substantial radial or vertical magnetic flux are unlikely to ever achieve a dynamically significant magnetic pressure, whether via the MRI or $\Omega$-dynamo. This statement appears to have recently gained additional support from the reported thermal collapse of simulated thin disks threaded only by toroidal magnetic fields \citep{Liska22}.

There are many plausible magnetic field topologies for accretion disks; here we have only considered four very simplified ones. It could be that in real astrophysical systems there may be a few preferred topologies. Recent EHT polarization measurements of M87 suggested an organized poloidal field component in the near-horizon region \citep{EHT1,EHT2}. \citet{Sadowski16b} gave an estimate for how strong such fields would need to be to stabilize the disks in particular X-ray binaries. Although he argued that it is reasonable for such field strengths to be provided by the companion star, he left open the question of how the fields might reach the inner accretion disk regions where thermal stability is a question. We, too, have dodged this important question for now.

One thing our work adds to this debate is that a net magnetic flux may not be necessary for thermal stability. In most accretion flows, the toroidal magnetic component will dominate due to the $\Omega$-dynamo, even in so-called MAD disks \citep{Begelman22}. This component can provide quite high magnetic pressure, even while supporting the MRI \citep{Wielgus15}. The question is really how this toroidal component sustains itself. This is where either a background vertical or extended radial field becomes crucial, as it will allow the toroidal component to be continually regenerated. One possible new requirement from our current work is that in order to achieve the strengths required to stabilize radiation-pressure-dominated disks, a radial field configuration must not be strongly affected by a midplane current sheet.

\section*{Acknowledgements}
We would like to thank Greg Salvesen, Wlodek Kluzniak and Mitch Begelman for useful feedback on this work. This work used the Extreme Science and Engineering Discovery Environment (XSEDE), which is supported by National Science Foundation grant number ACI-1548562. PCF gratefully acknowledges the support of the National Science Foundation through the following grants: AST1616185, PHY-1748958, and AST-1907850. This work was performed in part at the Aspen Center for Physics, which is supported by National Science Foundation grant PHY-1607611. BM acknowledges computing support from PROMETHEUS supercomputer in the PL-grid infrastructure in Poland. BM also acknowledges the funding support from LDRD grant at LANL. JA and AD extend their gratitude for the Summer Undergraduate Research with Faculty (SURF) grants they received from the office of Undergraduate Research and Creative Activities (URCA) at the College of Charleston.

\software{Cosmos++ \citep{Anninos05, Fragile12, Fragile14}}

\bibliography{refs}

\end{document}